\documentclass[journal,twoside,web]{ieeecolor}
\usepackage{tmi}

\usepackage{etoolbox}
\makeatletter
\@ifundefined{color@begingroup}%
{\let\color@begingroup\relax
   \let\color@endgroup\relax}{}%
\def\fix@ieeecolor@hbox#1{%
  \hbox{\color@begingroup#1\color@endgroup}}
\patchcmd\@makecaption{\hbox}{\fix@ieeecolor@hbox}{}{\FAILED}
\patchcmd\@makecaption{\hbox}{\fix@ieeecolor@hbox}{}{\FAILED}

\usepackage{cite}
\usepackage{amsmath,amssymb,amsfonts}
\usepackage{algorithmic}
\usepackage{textcomp}
\usepackage{mwe} % to get dummy images
\usepackage{multirow}
\usepackage{siunitx}
\usepackage{tikz}
\usepackage{colortbl} % in the preamble
\usepackage{pgfplots}
\usepackage{makecell}
\usepackage[colorlinks, citecolor=green]{hyperref}
\usepackage{adjustbox}
\usepackage{hhline}
\usepackage{booktabs}
\usepackage{svg}
\usepackage{arydshln}
\usepackage{graphicx,verbatim}
\usepackage{float}

\definecolor{lightgray}{gray}{0.9}

\def\BibTeX{{\rm B\kern-.05em{\sc i\kern-.025em b}\kern-.08em
    T\kern-.1667em\lower.7ex\hbox{E}\kern-.125emX}}
\begin{document}

\title{Latent Interpolation Learning Using Diffusion Models for Cardiac Volume Reconstruction}

\author{Niklas Bubeck, Suprosanna Shit, Chen Chen, Can Zhao, Pengfei Guo, Dong Yang, Georg Zitzlsberger, Daguang Xu, Bernhard Kainz, \textit{Senior Member, IEEE}, Daniel Rueckert, \textit{Fellow, IEEE}, Jiazhen Pan
\thanks{This research study was conducted retrospectively using human subject data made available in open access by the UK Biobank Resource under Application Number 87802. Ethical approval was not required as confirmed by the license attached with the open access data. This work is funded in part by the European Research Council (ERC) project Deep4MI (884622) and the Munich Center for Machine Learning.}
\thanks{N. Bubeck is with the School of Computation, Information and Technology, Technical University Munich, Germany. And also with the Munich Center for Machine Learning, Germany (e-mail:niklas.bubeck@tum.de).\\ 
\indent S. Shit is with the Department of Quantitative Biomedicine, University of Zurich, Switzerland. \\
\indent C. Chen is with the Institute of Biomedical Engineering, Department of Engineering Science, University of Oxford, UK. Also with Imperial College London, UK. And also with the University of Sheffield, UK.\\ 
\indent C. Zhao, P. Guo, D. Yang, G. Zitzlsberger and D. Xu are with NVIDIA.\\ 
\indent B. Kainz is with the Department of Computing, Imperial College London, UK. And also with the Department AIBE of Friedrich-Alexander-Universit\"at Erlangen-N\"urnberg, Germany. \\
\indent D. Rueckert is with the School of Computation, Information and Technology, Technical University Munich, Germany. Also with the Munich Center for Machine Learning, Germany. Also with the School of Medicine, Klinikum Rechts der Isar, Technical University of Munich, Germany. And also with the Department of Computing, Imperial College London, UK.\\
\indent Jiazhen Pan is with the School of Computation, Information and Technology, Technical University Munich, Germany. And also with the School of Medicine, Klinikum Rechts der Isar, Technical University of Munich, Germany.
}}

\maketitle

\begin{abstract}
Cardiac Magnetic Resonance (CMR) imaging is a critical tool for diagnosing and managing cardiovascular disease, yet its utility is often limited by the sparse acquisition of 2D short-axis slices, resulting in incomplete volumetric information. Accurate 3D reconstruction from these sparse slices is essential for comprehensive cardiac assessment, but existing methods face challenges, including reliance on predefined interpolation schemes (e.g., linear or spherical), computational inefficiency, and dependence on additional semantic inputs such as segmentation labels or motion data. To address these limitations, we propose a novel \textbf{Ca}rdiac \textbf{L}atent \textbf{I}nterpolation \textbf{D}iffusion (CaLID) framework that introduces three key innovations. First, we present a data-driven interpolation scheme based on diffusion models, which can capture complex, non-linear relationships between sparse slices and improves reconstruction accuracy. Second, we design a computationally efficient method that operates in the latent space and speeds up generative 3D whole-heart upsampling time by a factor of 24, reducing computational overhead compared to previous methods. Third, with only sparse 2D CMR images as input, our method achieves SOTA performance against baseline methods, eliminating the need for auxiliary input such as morphological guidance, thus simplifying workflows. We further extend our method to 2D+T data, enabling the effective modeling of spatiotemporal dynamics and ensuring temporal coherence. Extensive volumetric evaluations and downstream segmentation tasks demonstrate that CaLID achieves superior reconstruction quality and efficiency. By addressing the fundamental limitations of existing approaches, our framework advances the state of the art for spatio and spatiotemporal whole-heart reconstruction, offering a robust and clinically practical solution for cardiovascular imaging.
\end{abstract}


\section{Introduction}
\label{sec:introduction}
\IEEEPARstart{C}{ardiac} Magnetic Resonance (CMR) imaging plays a vital role in clinical diagnosis, treatment planning, and monitoring of cardiovascular diseases by providing detailed functional and anatomical information of the heart~\cite{bai2020population,zhang2024whole,wang2024screening,zhang2025towards}. To mitigate artifacts from respiratory and cardiac motion, scans are usually completed rapidly using breath-hold, balanced steady-state free precession (bSSFP) sequences with high in-plane resolution. The stringent time constraints require k-space undersampling and enormous effort is required to deliver high-quality CMR images~\cite{mun2013motion,zhao2019motion,chang2022deeprecon,pan2024reconstruction,pan2024unrolled}. As a result, clinicians typically acquire a stack of 2D short-axis (SAX) slices with substantial inter-slice gaps (8–10 mm) rather than a densely sampled 3D volume. Although this approach is more practical and time-efficient than dense 3D acquisition, it results in incomplete volumetric information and potential loss of critical cardiac structures. 

Alternatively, whole-heart acquisitions enable benefits such as simplified scan planning, free-breathing, and the retrospective reformatting of any cardiac view \cite{monney2015single,piccini2012respiratory,ogier2025free,moghari2018free, holtackers2025low}. However, these techniques remain limited due to significant trade-offs against image fidelity, including reduced blood-myocardium contrast, inferior image sharpness, and lower observer confidence compared to standard 2D cine. Furthermore, many of these approaches require the administration of contrast agents to improve quality.

Therefore, accurate 3D reconstruction from sparsely acquired 2D multislice stacks remains crucial for todays clinical applications, as it delivers comprehensive information on cardiac morphology and function under superior image fidelity, ultimately enabling more informed diagnostic and therapeutic decisions.

Despite notable advances in cardiac reconstruction, estimating patient-specific 3D volumes from sparsely sampled 2D CMR slices remains inherently ill-posed, compounded by the lack of dense 3D ground truth in routine clinical data. Recent cardiac reconstruction models ~\cite{chang2022deeprecon,jayakumar2023sadir,ye2023neural,qiao2025personalized}
have advanced cardiac analysis, yet they typically rely on mesh- or shape-based representations, supervised geometric priors, or dense volumetric supervision. These assumptions favor surface- or shape-level modeling but limit applicability to voxel-domain reconstruction from sparse slices, where interior intensity recovery and volumetric consistency must be inferred from limited observations.

To address this, prior approaches commonly solve the ill-posed nature by treating upsampling as an interpolation task between adjacent slices to infer missing anatomical structures. However, a key limitation of these methods is their \textbf{dependence on predefined interpolation schemes}, such as linear~\cite{stylegan,karras2020analyzing,preechakul2021diffusion} or spherical interpolation~\cite{he2023dmcvr,zheng2024noisediffusion}, which may fail to capture complex spatial variations and anatomical detail. These fixed strategies impose strong assumptions about the relationship between adjacent slices. However, these schemes with human-imposed interpolation rules are often over simplistic, failing to fully capture the heart's intricate structural variations and temporal dynamics. For example,  image-space interpolation (e.g., bilinear or trilinear) schemes struggle to capture complex cardiac anatomical variations, while latent-space approaches are often limited to account for the inherent high non-linearity of the encoded representations. Another significant challenge is \textbf{computational inefficiency}. Methods operating in the image domain often require extensive optimization steps~\cite{chang2022deeprecon,stolt2023nisf} or long diffusion sampling chains~\cite{he2023dmcvr}, making them impractical for clinical use. Additionally, many current approaches depend on \textbf{extra semantic/morphological information as inputs}, such as segmentation labels or cardiac motion data~\cite{chang2022deeprecon,wang2021joint,frakes2008new,leng2013medical,meng2022mulvimotion}, which increases both the complexity of the pipeline and the data annotation burden. Furthermore, existing approaches such as \cite{chang2022deeprecon,he2023dmcvr,bubeck2025reconstruct} are either restricted to \textbf{purely 2D spatial interpolation}, which prevents them from generating temporally consistent \(2D+T\) sequences, or they model \textbf{temporal progression only through conditioning} \cite{biller2026tumorflow}, rather than explicitly learning the underlying spatiotemporal structure of the data. This lack of temporal modeling leads to discontinuities and incoherent morphology across frames, limiting the applicability of such methods in dynamic cardiac imaging. These limitations underscore the need for a more adaptive, efficient and clinically feasible solution that can directly leverage the sparse CMR data without requiring additional inputs.

In this work, we present a novel \textbf{Ca}rdiac \textbf{L}atent \textbf{I}nterpolation \textbf{D}iffusion (CaLID) framework mitigating these challenges via three key innovations:
\begin{enumerate}
\item \textit{\textbf{Data-driven latent space interpolation learning}}: Rather than imposing predefined interpolation rules, CaLID learns optimal interpolation strategies directly from the data in the latent space. This allows the network to discover complex, non-linear relationships between sparse slices, leading to more accurate reconstruction of cardiac anatomical structures. By operating in the latent space, our model can better capture the intricate variations in cardiac anatomy that fixed interpolation schemes often miss.
\item \textit{\textbf{Computationally efficient}}: CaLID achieves remarkable efficiency through a carefully designed training pipeline and accurate in-distribution image interpolation. By operating in the latent space and requiring only 8 diffusion steps during 3D and 3D+T (from 2D and 2D+T) whole-heart upsampling, we dramatically reduce computational overhead compared to image-domain methods that often require hundreds of steps. This efficiency makes our approach practical for clinical deployment while maintaining high reconstruction quality.
\item \textit{\textbf{Superior performance with minimal input requirements}}: Unlike existing methods that rely on additional semantic priors such as segmentation labels, CaLID achieves state-of-the-art reconstruction using only sparse 2D or 2D+T CMR images, without the need for any auxiliary inputs or annotations. We demonstrate superior performance through comprehensive evaluation and downstream segmentation tasks. The ability to achieve such results without auxiliary input not only simplifies the clinical workflow but also makes our method more widely applicable across different clinical settings.
\end{enumerate}

\section{Background and Related Work}
\noindent Deep generative models have achieved remarkable progress in high-quality image synthesis, with generative adversarial networks (GANs) historically setting the standard for sample fidelity. However, GANs are often challenged by unstable training dynamics and mode collapse~\cite{bau2019seeing}, prompting exploration of alternative likelihood-based approaches such as variational autoencoders~\cite{kingma2013auto}, autoregressive models~\cite{esser2021taming}, and normalizing flows~\cite{rezende2015variational}. Among these, diffusion and score-based models have emerged as a promising class, capable of generating samples without adversarial training. Specifically, Denoising Diffusion Probabilistic Models DDPMs)~\cite{ho2020denoising,song2020score,dhariwal2021diffusion,ho2022classifier} map between a Gaussian prior $\mathcal{N}(\mathbf{0}, \mathbf{I})$ and targeted data distribution by learning to successively reverse a predefined Markovian forward process. The forward process gradually corrupts a clean image $x_0$ with Gaussian noise over $T$ timesteps, generating intermediate noisy states $x_t$ using the marginals:
\begin{equation}\label{equation1}
    q(x_t \mid x_0) = \mathcal{N}(\sqrt{\alpha_t}x_0, (1 - \alpha_t)I),
\end{equation}
where $\alpha_t$ controls the noise schedule. To learn the mapping to the target distribution, a neural network $\epsilon_\theta(x_t, t)$ is trained to predict the actual noise $\epsilon_t$ added for each timestep $t$, minimizing the objective function $\| \epsilon_\theta(x_t, t) - \epsilon_t \|^2$. By applying the model sequentially over multiple steps, the process can be reversed again using the following formulation: 
\begin{align}\label{equation2}
    x_{t-1} =\ & \sqrt{\alpha_{t-1}} \left( \frac{x_t - \sqrt{1 - \alpha_t} \epsilon_\theta(x_t, t)}{\sqrt{\alpha_t}} \right) \nonumber \\
    & + \sqrt{1 - \alpha_{t-1} - \sigma^2_{t}} \cdot  \epsilon_\theta(x_t, t) + \sigma_t\epsilon_t,
\end{align}
where $\sigma_t\epsilon_t$ is some random noise weighted by $\sigma_t$ that depends on the noising schedule.

\noindent Although DDPMs produce high-quality samples, their stochastic sampling process typically requires hundreds of steps, making them computationally intensive. Denoising Diffusion Implicit Models (DDIMs)~\cite{song2020denoising} address this limitation by introducing a deterministic reverse process that generalizes the diffusion framework through non-Markovian forward processes while retaining the original training objective and forward process marginals as in Eq.~\ref{equation1}. The reverse process is described as:  
\begin{align}\label{equation3}
    x_{t-1} =\ & \sqrt{\alpha_{t-1}} \left( \frac{x_t - \sqrt{1 - \alpha_t} \epsilon_\theta(x_t, t, c^*)}{\sqrt{\alpha_t}} \right) \nonumber \\
    & + \sqrt{1 - \alpha_{t-1}} \cdot \epsilon_\theta(x_t, t, c^*)
\end{align}
where $c^*$ is an optional condition. By setting $\sigma=0$ the reverse process becomes deterministic, which makes it possible to obtain a noisy latent representation $x_T$ representing a noisy encoding of a given image $x_0$. This deterministic encoding into noisy latent representations is also known as diffusion inversion and enables manipulation over the generated image. However, recent work~\cite{preechakul2021diffusion} has shown that the terminal noisy latent representation $x_T$ produced by DDIM inversion lacks meaningful semantics, thus limiting its usefulness for interpolation, reconstruction, or downstream representation learning.
Diffusion Autoencoders (DiffAE)~\cite{preechakul2021diffusion} resolve this by disentangling semantics and stochasticity. They introduce a latent code pair $z_{\text{sem}}$ for semantic content, $x_T$ for stochastic variation, and use it ($c^*=z_{sem}$ ) to condition a diffusion model to reconstruct $x_0$. This approach allows for more meaningful semantic interpolation in $z_{\text{sem}}$ and provides increased control over generated content. Manipulation is then performed by identifying semantic directions within the latent space and applying controlled perturbations along these directions.

For domain-specific applications such as cardiac MRI reconstruction, DMCVR~\cite{he2023dmcvr} enhances diffusion-based methods through a morphology-guided framework. Each 2D cine-MRI slice is decomposed into three latent components: a global semantic code, a regional morphology code capturing fine-grained anatomical detail, and a noisy representation obtained via DDIM inversion. These components are interpolated (most commonly using linear or spherical interpolation) and subsequently decoded into intermediate 2D slices that are stacked to form the final 3D volume. Thereby, spherical interpolation operates by following geodesic paths on a hypersphere between latent codes, preserving their norms to promote smooth transitions. However, such static schemes remain agnostic to the data distribution and do not preserve anatomical plausibility when interpolating highly non-linear latent representations.

\begin{figure*}[t]
    \centering
\includegraphics[width=0.98\textwidth]{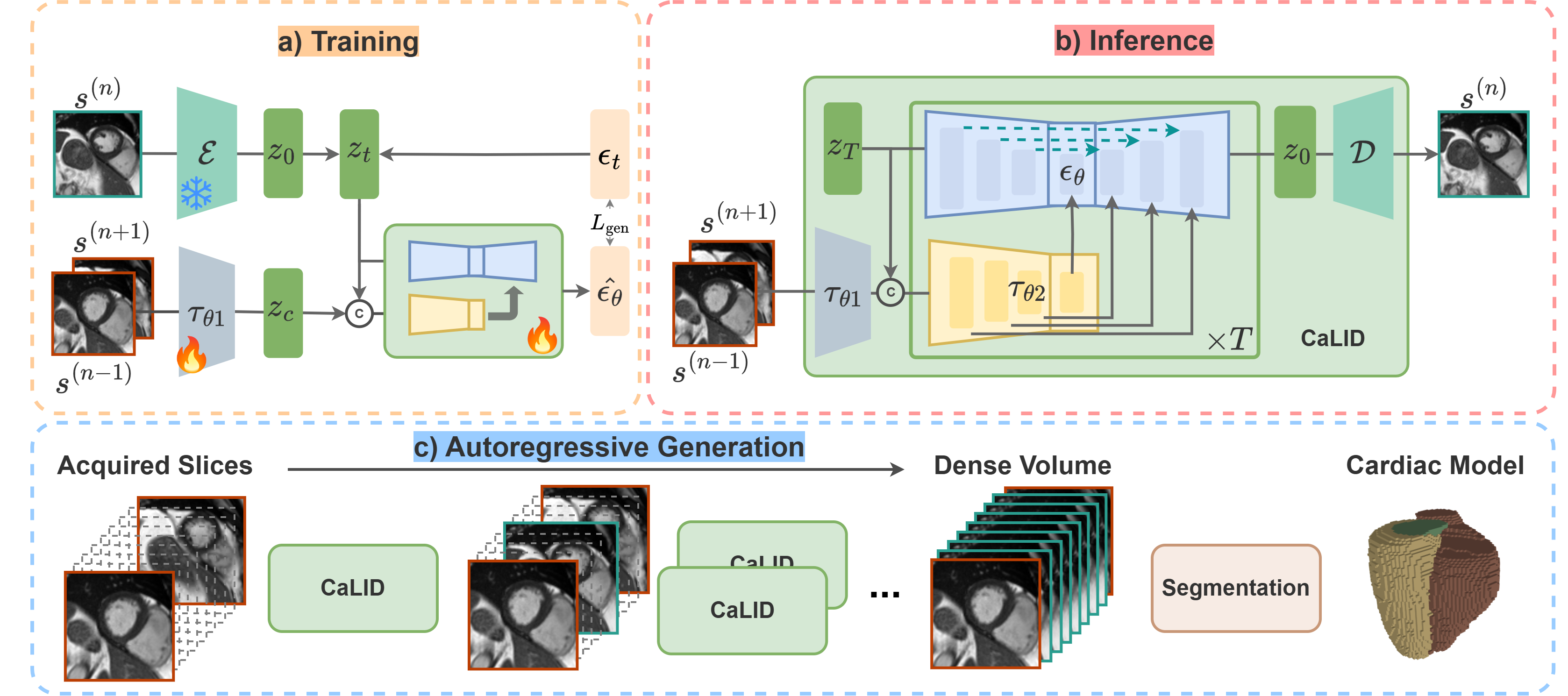}
    \caption{\textbf{\textit{Our proposed CaLID framework.}} 
    (a) End-to-end training strategy incorporating a two-part conditioning network ($\tau_{\theta 1}$, $\tau_{\theta 2}$) and U-Net denoiser ($\epsilon_{\theta}$), optimized jointly for data-driven interpolation. (b) Inference architecture: generation of intermediate slice $s^{(n)}$ from neighboring slices $s^{(n-1)}$ and $s^{(n+1)}$ using a latent diffusion model. (c) Autoregressive application of CaLID to generate missing slices, enabling whole-heart segmentation. The illustrated example shows the 2D case, but the same framework is also applicable to spatiotemporal 2d+T data.}
    \label{fig:main_figure}
\end{figure*}

Consequently, both DiffAE and DMCVR treat interpolation as a secondary task: trained primarily for reconstruction, they are later adapted for interpolation by manipulating noisy latent variables \(x_T\) and semantic features \(z_{\text{sem}}\) through fixed \emph{post hoc} schemes. This mismatch between training objective and downstream application inherently constrains their performance. When the latent space lacks smoothness or semantic structure, interpolation trajectories may produce incoherent or anatomically implausible outputs. In addition, obtaining latent codes typically requires computationally expensive inversion, while diffusion sampling in the high-dimensional pixel space further increases the computational burden.

Another key limitation arises from exposure bias in autoregressive generation \cite{ning2023elucidating,ning2023input}. During training, models are conditioned on ground-truth contexts, whereas at inference time they rely on their own previously generated outputs. Consequently, small local errors can accumulate over time, leading to drift, blurring, or anatomical artifacts. This effect is particularly pronounced when fixed latent mixing strategies are used, as intermediate states are not explicitly optimized to maintain anatomical consistency.

Our approach addresses these limitations by explicitly training the model for interpolation rather than adapting a reconstruction-based objective. By incorporating interpolation directly into the training objective, the learned dynamics remain aligned with the underlying data distribution, promoting anatomical coherence while avoiding costly image-space inversion. Furthermore, by constraining interpolation to a dense learned manifold and enforcing strong local conditioning at each recursive step, the proposed method stabilizes generation trajectories, mitigates error propagation, and enables efficient, patient-consistent generative modeling.

\section{Methods}
\noindent Our proposed framework illustrated in Figure~\ref{fig:main_figure} implements a bisectional-based autoregressive scheme for volumetric reconstruction. This architecture progressively synthesizes intermediate SAX slices through iterative refinement, enabling systematic reconstruction of a high-resolution 3D volumetric representation from sparse 2D acquisitions (Section~\ref{inference}). Inspired by Latent Diffusion Models (LDMs)~\cite{rombach2022high}, the core component is a latent diffusion model conditioned on adjacent anatomical context through paired neighboring slices $s^{(n-1)}$ and $s^{(n+1)}$. This formulation enables data-driven interpolation of the intermediate slice $s^{(n)}$ while preserving anatomical consistency, with complete training dynamics specified in Section~\ref{training}. Furthermore, our framework extends to spatiotemporal interpolation showing increased temporal coherence.

\subsection{Training}\label{training}

% \noindent Our method employs a Variational Autoencoder (VAE), consisting of an encoder \(\mathcal{E}\) and a decoder \(\mathcal{D}\), specifically tailored for cardiac MRI data. Inspired by~\cite{rombach2022high}, we train the VAE from scratch using CMR SAX images, finding that this domain-specific training preserves cardiac anatomical structures more effectively than general-purpose VAEs trained on natural images. The encoder \(\mathcal{E}\) maps an input slice \(s^{(n)}\) to a continuous latent representation \(z_0 = \mathcal{E}(s^{(n)})\), approximating the posterior distribution \(q_\phi(z|x)\) with \(z \sim \mathcal{N}(\mu_z, \Sigma_z)\). The decoder \(\mathcal{D}\) reconstructs the image by estimating the likelihood \(p_\theta(x|z)\). We train the VAE by maximizing the Evidence Lower Bound (ELBO):

% \begin{align}
% \mathcal{L}(\theta, \phi; x) = -\mathrm{D_{KL}}\big(q_\phi(z|x) \,\|\, p_\theta(z)\big) \nonumber \\
% + \mathbb{E}_{q_\phi(z|x)}\big[\log p_\theta(x|z)\big]
% \end{align}

% \noindent where \(\mathrm{D_{KL}}\) denotes the Kullback-Leibler divergence and the prior \(p(z) = \mathcal{N}(0, I)\) is a standard Gaussian by design. After pretraining, both \(\mathcal{E}\) and \(\mathcal{D}\) are frozen, and their latent space is leveraged as the foundation for a diffusion-based generative model trained as a second stage.

\noindent Our method employs a Variational Autoencoder (VAE), consisting of an encoder \(\mathcal{E}\) and a decoder \(\mathcal{D}\), specifically tailored for cardiac MRI data. Inspired by~\cite{rombach2022high}, we train the VAE from scratch using CMR SAX images, finding that this domain-specific training preserves cardiac anatomical structures more effectively than general-purpose VAEs trained on natural images. The encoder \(\mathcal{E}\) maps an input slice \(s^{(n)}\) to a continuous latent representation \(z_0 = \mathcal{E}(s^{(n)})\), approximating the posterior distribution \(q_\phi(z|x)\) with \(z \sim \mathcal{N}(\mu_z, \Sigma_z)\). The decoder \(\mathcal{D}\) reconstructs the image by estimating the likelihood \(p_\theta(x|z)\). 
% We train the VAE by maximizing the Evidence Lower Bound (ELBO):

% \begin{align}
% \mathcal{L}(\theta, \phi; x) = -\mathrm{D_{KL}}\big(q_\phi(z|x) \,\|\, p_\theta(z)\big) \nonumber \\
% + \mathbb{E}_{q_\phi(z|x)}\big[\log p_\theta(x|z)\big]
% \end{align}

% \noindent where \(\mathrm{D_{KL}}\) denotes the Kullback-Leibler divergence and the prior \(p(z) = \mathcal{N}(0, I)\) is a standard Gaussian by design.

\noindent Optimizing the Evidence Lower Bound (ELBO) alone using the Kullback-Leibler divergence \(\mathrm{D_{KL}}\) often leads to overly smooth reconstructions and sparsely populated latent regions, which is undesirable for later interpolation. Therefore, following~\cite{rombach2022high}, we augment the training with perceptual and adversarial constraints.
In our implementation, \(F\) corresponds to the LPIPS AlexNet backbone and \(l\) indexes its convolutional feature blocks:
\begin{align}
\mathcal{L}_{perc} =
\sum_l \|F_l(x) - F_l(\hat{x})\|_2^2 ,
\end{align}
which enforces similarity of anatomical structures beyond pixel-wise alignment.
Additionally, we employ a PatchGAN critic \(\mathcal{C}_\psi\) and optimize the adversarial objectives
\begin{align}
\mathcal{L}_{GAN}^{G} &= -\mathbb{E}_{z \sim q_\phi(z|x)}[\log \mathcal{C}_\psi(\hat{x})], \\
\mathcal{L}_{GAN}^{C} &= -\mathbb{E}_{x}[\log \mathcal{C}_\psi(x)]
-\mathbb{E}_{z \sim q_\phi(z|x)}[\log (1-\mathcal{C}_\psi(\hat{x}))].
\end{align}
The complete autoencoder optimization objective becomes
\begin{equation}
\begin{aligned}
\mathcal{L}_{VAE} &=
\lambda_{rec}\|x-\hat{x}\|_1
+ \lambda_{perc}\mathcal{L}_{perc} \\
&\quad + \lambda_{KL}\mathrm{D_{KL}}\big(q_\phi(z|x)\|p(z)\big)
+ \lambda_{adv}\mathcal{L}_{GAN}^{G}.
\end{aligned}
\end{equation}
where $\theta$, $\phi$ and $\psi$ are learnable network parameters, and $\lambda$ is a weighting factor for each loss component. 
The KL term enforces global continuity of the latent distribution, the reconstruction term preserves pixel fidelity, the perceptual term maintains perceptual reconstruction, and the adversarial term discourages blurry averages. Together, these constraints produce a smooth and densely populated latent manifold, which is crucial for the later interpolation.

\noindent After pretraining, both \(\mathcal{E}\) and \(\mathcal{D}\) are frozen, and their latent space is leveraged as the foundation for a diffusion-based generative model trained as a second stage.

Specifically we train a U-Net based denoiser \(\epsilon_{\theta}\) to predict the noise \(\epsilon_t\) added to the latent code \(z_0\) at each diffusion timestep \(t\). Training minimizes the generative loss, defined as the expected squared error between the predicted noise \(\hat{\epsilon}_{\theta}\) and the true noise \(\epsilon_t\), conditioned on adjacent slices:

\begin{align}
L_{\mathrm{gen}}(z) =\ & \mathbb{E}_{t, z_0, \epsilon} \Bigl[
\bigl\| \epsilon_{\theta}\bigl(z_t, t, 
\tau_{\theta 2}(\tau_{\theta 1}(s^{(n-1)}, s^{(n+1)}), z_t) \bigr) \nonumber \\
& - \epsilon_t \bigr\|_2^2 \Bigr]
\end{align}

\noindent where \(\tau_{\theta1}\) and \(\tau_{\theta2}\)  denote to our conditioning mechanism.

\noindent\textbf{\textit{Conditioning Mechanism.}}
To facilitate anatomically consistent interpolation across slices, we introduce a two-stage conditioning module composed of $\tau_{\theta 1}$ and $\tau_{\theta 2}$. The first component, $\tau_{\theta 1}$, encodes contextual information by taking as input the adjacent slices $[s^{(n-1)}, s^{(n+1)}]$ and projecting them into a latent conditional embedding $z_c$. Importantly, we do not directly reuse the pretrained VAE encoder for this purpose. While the VAE latent space is intentionally regularized to be smooth and well-structured for interpolation, conditioning requires expressive predictive capacity rather than strict latent continuity. Using the VAE encoder would bias the conditional signal towards reconstruction semantics and overly constrain the conditioning process. Instead, $\tau_{\theta 1}$ learns a dedicated conditional representation that can capture variability between neighboring slices without being restricted by the interpolation-oriented latent manifold.
This is achieved using a convolutional network architecture with a terminal zero-convolution layer, following best practices from~\cite{lin2024diffbir} to encourage faster and more stable convergence. The second component, $\tau_{\theta 2}$, is inspired by the architecture of ControlNet~\cite{zhang2023adding} and serves to inject the conditional information $z_c$ into the denoising process. Unlike traditional ControlNet implementations, which employ a two-stage training procedure with frozen backbones, we adopt a single-stage joint training strategy. Specifically, we simultaneously optimize $\epsilon_{\theta}$, $\tau_{\theta 1}$, and $\tau_{\theta 2}$, encouraging tighter integration between conditioning and denoising and simplifying the training pipeline.

\noindent\textbf{\textit{Spatiotemporal Flexibility.}}
To enable fast and temporally coherent reconstruction of dynamic cine MRI sequences, we extend our model to handle spatiotemporal data by interpreting time as an additional spatial dimension. Consequently, we replace all 2D convolutions in the encoder $\mathcal{E}$, decoder $\mathcal{D}$, denoiser $\epsilon_{\theta}$, and conditioning modules $\tau_{\theta 1}$, $\tau_{\theta 2}$ with their 3D counterparts. This simple architectural change enables joint modeling of spatial and temporal features, allowing our framework to seamlessly generalize to spatiotemporal data. As a result, the proposed method offers a unified, scalable solution for diverse CMR reconstruction and generation tasks across static and dynamic settings.

\noindent\textbf{\textit{Difference to DiffAE and DMCVR.}} While DiffAE~\cite{preechakul2021diffusion} and DMCVR~\cite{he2023dmcvr} also employ diffusion-based interpolation, our method differs in the following key aspects:

\noindent\textit{1) Latent Space Interpolation} DiffAE and DMCVR perform fixed interpolation in the noise space of the diffusion process (i.e., interpolating between $x_T$ representations). In contrast, our method learns data-driven trajectories using the conditioning networks $\tau_{\theta 1}$ and $\tau_{\theta 2}$ in the autoencoder’s latent space (i.e., $z_0 = \mathcal{E}(s^{(n)})$), which is semantically structured and better aligned with anatomical content. This leads to more anatomically meaningful interpolations and limits drift due to exposure bias.

\noindent\textit{2) Conditioning Granularity.} Both DiffAE and DMCVR inject semantic information at a single bottleneck layer, limiting their ability to capture high-resolution spatial details. Our architecture supports multi-scale conditioning, introducing context at several stages of the U-Net denoiser, which improves both global structure and fine anatomical accuracy (see Figure~\ref{fig:pred_all} and~\ref{fig:sax_to_lax}). 

\noindent\textit{3) Requirements and dependencies.} DMCVR requires segmentation labels during training, increasing the need for annotated data. In contrast, our framework is fully self-supervised, operating directly on raw SAX slices, which simplifies preprocessing and enhances scalability in clinical applications.

\subsection{Inference}\label{inference}

\noindent\textbf{\textit{Autoregressive Generation.}} Following training, our model supports the synthesis of intermediate slices by conditioning on neighboring slices using an autoregressive strategy. To improve efficiency at inference time, we adopt the DDIM sampling method for the latent space reducing the number of diffusion steps while maintaining image fidelity, which is defined as: 
\begin{align}\label{equation4}
    z_{t-1} =\ & \sqrt{\alpha_{t-1}} \left( \frac{z_t - \sqrt{1 - \alpha_t} \epsilon_\theta(z_t, t, c^*)}{\sqrt{\alpha_t}} \right) \nonumber \\
    & + \sqrt{1 - \alpha_{t-1}} \cdot \epsilon_\theta(z_t, t, c^*) \quad .
\end{align}
The conditioning here is the output of our conditioning mechanism, thus $c^*=\tau_{\theta 2}(\tau_{\theta 1}(s^{(n-1)}, s^{(n+1)}), z_t)$.
Although the model is trained to reconstruct discrete slices $s^{(n)}$ from adjacent inputs $\bigl[s^{(n-1)}, s^{(n+1)}\bigr]$, it generalizes to non-integer positions during inference. For example, a slice at position $s^{(n+0.5)}$ can be generated by conditioning on $\bigl[s^{(n)}, s^{(n+1)}\bigr]$. This forms the basis for a bisection-style autoregressive interpolation: once $s^{(n+0.5)}$ is generated, we can recursively synthesize finer interpolations such as $s^{(n+0.25)}$ by conditioning on $s^{(n)}$ and $s^{(n+0.5)}$, and so on. Repeating this process allows for flexible and arbitrarily high-resolution upsampling of the 2D slice stack.

\begin{table*}[t]
\centering
\caption{Quantitative comparison of slice interpolation for $s^n$ from adjacent slices $s^{n-1}$ and $s^{n+1}$ using 2D and 2D+T models. Our methods CaLID and CaLID\textsubscript{+} using initialization refinement consistently outperform prior methods across all metrics for spatial and spatiotemporal slice interpolation, while providing a tunable trade-off between fidelity and inference speed.}
\renewcommand{\arraystretch}{1.1} 
\resizebox{0.9\textwidth}{!}{%
\begin{tabular}{c|l|ccccc}
  \cline{1-7}
  \textbf{Dim.} & \textbf{Model} &
  \textbf{PSNR} $\uparrow$ &
  \textbf{SSIM} $\uparrow$ &
  \textbf{LPIPS\textsubscript{alex}} $\downarrow$ &
  \textbf{rFID} $\downarrow$ &
  \textbf{Gen. Time\textsubscript{128}} $\downarrow$ \\
  \cline{1-7}
  \multirow{6}{*}{\centering 2D} 
    & Bilinear Pixel
      & $19.795_{\pm 1.55}$
      & $0.535_{\pm 0.05}$
      & $0.149_{\pm 0.02}$
      & $84.050$
      & -- \\
    & Bilinear Latent
      & $20.397_{\pm 1.64}$
      & $0.586_{\pm 0.05}$
      & $0.172_{\pm 0.03}$
      & $92.685$
      & -- \\
    & DiffAE~\cite{preechakul2021diffusion}
      & $20.741_{\pm 1.44}$
      & $0.597_{\pm 0.06}$
      & $0.127_{\pm 0.02}$
      & $88.111$
      & \underline{7.875s} \\
    & DMCVR~\cite{he2023dmcvr}
      & $21.023_{\pm 1.72}$
      & $0.631_{\pm 0.05}$
      & $0.124_{\pm 0.03}$
      & $87.995$
      & 8.966s  \\
    \cdashline{2-7}
    & CaLID
      & $\underline{22.748_{\pm 2.14}}$
      & $\underline{0.755_{\pm 0.07}}$
      & $\underline{0.097_{\pm 0.02}}$
      & $\underline{55.585}$
      & \textbf{3.025s} \\
    & CaLID\textsubscript{+}
      & $\boldsymbol{24.160_{\pm 2.35}}$
      & $\boldsymbol{0.778_{\pm 0.07}}$
      & $\boldsymbol{0.086_{\pm 0.02}}$
      & $\boldsymbol{31.508}$
      & 8.942s \\
  \cline{1-7}
  \multirow{2}{*}{\centering 2D+T} 
    & CaLID
      & $22.246_{\pm 1.91}$
      & $0.706_{\pm 0.07}$
      & $0.125_{\pm 0.02}$
      & $55.489$
      & 13.480s \\
    & CaLID\textsubscript{+}
      & $23.557_{\pm 2.11}$
      & $0.757_{\pm 0.06}$
      & $0.116_{\pm 0.02}$
      & $53.928$
      & 39.823s \\
  \cline{1-7}
\end{tabular}%
}
\label{table:metrics}
\end{table*}

\noindent\textbf{\textit{Initialization Refinement.}}
In standard diffusion-based inference, latent variables are typically initialized with pure Gaussian noise. While effective in many settings, such random initialization may lead to slower convergence or suboptimal outputs, especially in structured domains like medical imaging. To address this, we introduce an \textbf{optional} test-time refinement strategy based on diffusion inversion interpolation, similar in spirit to DiffAE and DMCVR. Specifically, we initialize the latent variable $z^n_T$ using a spherical linear interpolation (slerp)~\cite{shoemake1985animating,zheng2024noisediffusion} between the noisy latent representations of the two neighboring slices, $z^{(n-1)}_T$ and $z^{(n+1)}_T$. The slerp with interpolation coefficient $\gamma \in [0, 1]$ is defined as:
\begin{align}
z^n_T = \frac{\sin((1 - \gamma)\theta)}{\sin \theta} z^{(n-1)}_T + \frac{\sin(\gamma \theta)}{\sin \theta} z^{(n+1)}_T,
\end{align}

where $\theta$ defines the angular distance between the two noisy latent vectors measured as the arccosine of their normalized dot product. 

\begin{align}
\theta = \arccos\left(\frac{{z^{(n-1)}_T}^\top z^{(n+1)}_T}{\|z^{(n-1)}_T\|\|z^{(n+1)}_T\|}\right)
\end{align}
For uniformly distributed data, we set $\gamma=0.5$, which corresponds to interpolation at the midpoint of the spherical arc. This parameter can also be adjusted to compensate for non-uniform data. The initialization provides a more informed starting point for the denoising process, bringing the initialization closer to the data manifold and facilitating higher-quality reconstructions. Importantly, this initialization is not used during training, allowing the model to learn a free-form interpolation without relying on fixed trajectories. At inference time, however, it acts as a lightweight enhancement that improves generation fidelity without altering the model architecture or training procedure. We refer to our model with this refinement as $\text{CaLID}_{+}$. This way, we provide a flexible methodology that allows us to trade off between inference speed, computational efficiency, and reconstruction quality.

\noindent\textbf{\textit{Downstream Utility.}} Cardiac SAX upsampling enables downstream tasks by generating comprehensive cardiac models and clinical phenotypes from sparse input data. Following the upsampling, downstream models can be applied to extract key structures like left ventricle cavity (LVC), right ventricle cavity (RVC), and left ventricle myocardium (LVM), enabling cardiac modeling as well as phenotype estimation but also prediction directly from the upsampled stack.

While distinct from our core upsampling method, we demonstrate this utility by training a 2D MedFormer model~\cite{gao2022data} and apply it to our dense reconstructions. Additionally, we train it on real (non-augmented) data and use it as a qualitative out-of-distribution (OOD) measure (Fig.~\ref{fig:3d_recon}). Furthermore, a phenotype regression task confirms that upsampling maintains competitive phenotype prediction accuracy even with fewer input slices (Table~\ref{table:phenotype_comp_transposed}).

\section{Dataset and Experiments}
\noindent\textbf{\textit{Dataset.}}
The proposed method and baseline comparisons were implemented using cardiac MR short-axis (SAX) images from the UK BioBank dataset~\cite{petersen2015uk}, featuring voxel spacing of $1.8 \times 1.8 mm$ (axial) and $8 mm$ (longitudinal)
The dataset comprised 11,360 training and 1,420 testing subjects, with 50 temporal frames per subject (every 5th frame selected), during training we randomly select a slice, time frame and apply random flips as data augmentation. All SAX images were preprocessed via center-cropping to $128 \times 128 $ pixel cardiac regions. Corresponding long-axis (LAX) images served as the evaluation benchmark for interpolation performance. For the 2D+T setting, we uniformly subsample 32 frames to reduce computational load while maintaining temporal coverage across the cardiac cycle. This ensures the model learns dynamics representative of both systolic and diastolic phases.

\noindent\textbf{\textit{Implementation Details.}}
Both 2D and 2D+T VAEs adopt a design similar to that proposed by~\cite{rombach2022high}, utilizing a downsampling factor of $f=4$ to compress spatial dimensions. The VAE architectures do not employ attention mechanisms, and instead rely on convolutional operations. 
The networks $\tau_{\theta1}$ and $\tau_{\theta2}$ are implemented as U-Net encoder structures. While $\tau_{\theta1}$ is a lightweight encoder that performs hierarchical downsampling using convolutional ResBlocks without any attention mechanisms, $\tau_{\theta2}$ adopts a more expressive design incorporating time-embedded ResBlocks followed by self-attention modules at each resolution scale. The feature maps produced by $\tau_{\theta2}$ are injected into the main denoising network $\epsilon_{\theta}$ through multi-scale feature addition, enabling fine-grained conditioning during the denoising process.

\noindent\textbf{\textit{Training Details.}} All generative models underwent identical optimization procedures employing the Adam optimizer~\cite{kingma2014adam}, with a linear learning rate warmup over 50 epochs followed by a fixed learning rate of $1 \times 10^{-4}$. Training was performed on 2 NVIDIA A100 GPUs, using a batch size of 64 for the 2D setting and 16 for the 2D+T setting, and continued until convergence. Model performance was evaluated using an exponential moving average (EMA) of parameters with a decay rate of 0.999 to stabilize inference outcomes.

\noindent\textbf{\textit{2D Slice Generation.}}
To evaluate our method's performance, we conduct a comprehensive comparison against diffusion-based state-of-the-art approaches including DiffAE~\cite{preechakul2021diffusion}, DMCVR~\cite{he2023dmcvr}, as well as classical bilinear interpolation for pixel and latent space. Since ground-truth data for fractional slice positions is unavailable, we instead evaluate models on interpolating known integer-positioned slices. Specifically, given the adjacent slices $s^{(n-1)}$ and $s^{(n+1)}$, we task the model with predicting the intermediate slice $s^{(n)}$ and compare it against the ground-truth at hand. Our assessment employs both pixel-level fidelity metrics (PSNR), perceptual similarity measures (SSIM, LPIPS) and feature distances like the reconstruction Frechet Inception Distance (rFID) to gauge the quality of generated images. 

To better understand the \textit{computational implications}, we also investigate the trade-off between sampling efficiency and output quality. Specifically, we evaluate the generative performance of each method across varying numbers of diffusion steps, highlighting how our model maintains high fidelity even with fewer sampling iterations and measure the generation time of each model when using 128 diffusion steps. 

Furthermore, we evaluate the \textit{morphological plausibility} of the generated slices through a downstream segmentation task. Specifically, we apply the MedFormer~\cite{gao2022data} model to the predicted intermediate slice $s^{(n)}$ and compare the resulting segmentation mask against the ground-truth. This allows us to assess not only visual fidelity but also anatomical correctness. We report multiple evaluation metrics, including the Dice coefficient for region overlap, Hausdorff Distance (HD) for boundary extremity, Average Surface Distance (ASD), and Average Symmetric Surface Distance (ASSD) to quantify segmentation accuracy along structure boundaries.

To assess clinical relevance, we train three ResNet-50 \cite{he2016deep} models for \textit{downstream phenotype prediction} using fully acquired end-diastolic (ED) and end-systolic (ES) multi-slice cardiac volumes as reference data. The networks estimate left and right ventricular end-diastolic volumes (LVEDV, RVEDV) from ED volumes and left and right ventricular end-systolic volumes (LVESV, RVESV) from ES volumes. For the temporal setting, we additionally estimate left and right ventricular ejection fraction (LVEF, RVEF). 
To evaluate the effect of slice upsampling, ground-truth volumes are artificially sub-sampled by retaining every second slice and subsequently reconstructed to full resolution using our method and competing baselines. The trained regression models are applied to the reconstructed volumes, and the resulting prediction errors are compared with those obtained from the original fully sampled volumes, simulating a reduced slice acquisition protocol.
% During training, the interpolation model observes only adjacent slices $s^{(n-1)}$ and $s^{(n+1)}$ corresponding to an interpolation distance of $d=16\,\mathrm{mm}$.}

To assess \textit{robustness to larger inter-slice gaps}, we additionally evaluate interpolation at distances of $32\,\mathrm{mm}$ and $48\,\mathrm{mm}$ using slices $s^{(n\pm2)}$ and $s^{(n\pm3)}$, respectively. Furthermore, we investigate interpolation consistency by conditioning the model on asymmetric neighboring slices such that the predicted slice is not centered between the inputs, which we quantify using the uniformity factor $\gamma$.

\noindent\textbf{\textit{3D Volume Reconstruction.}}
We evaluate 3D reconstruction primarily through \emph{qualitative analysis}. Although native 3D cine could serve as a volumetric reference, a direct quantitative comparison is confounded given differences in acquisition protocol, effective resolution, motion characteristics, and contrast. Consequently, pixel- or feature-wise comparisons would largely reflect inter-sequence differences rather than interpolation fidelity.

\begin{figure*}[t]
    \centering
    \includegraphics[width=0.98\textwidth]{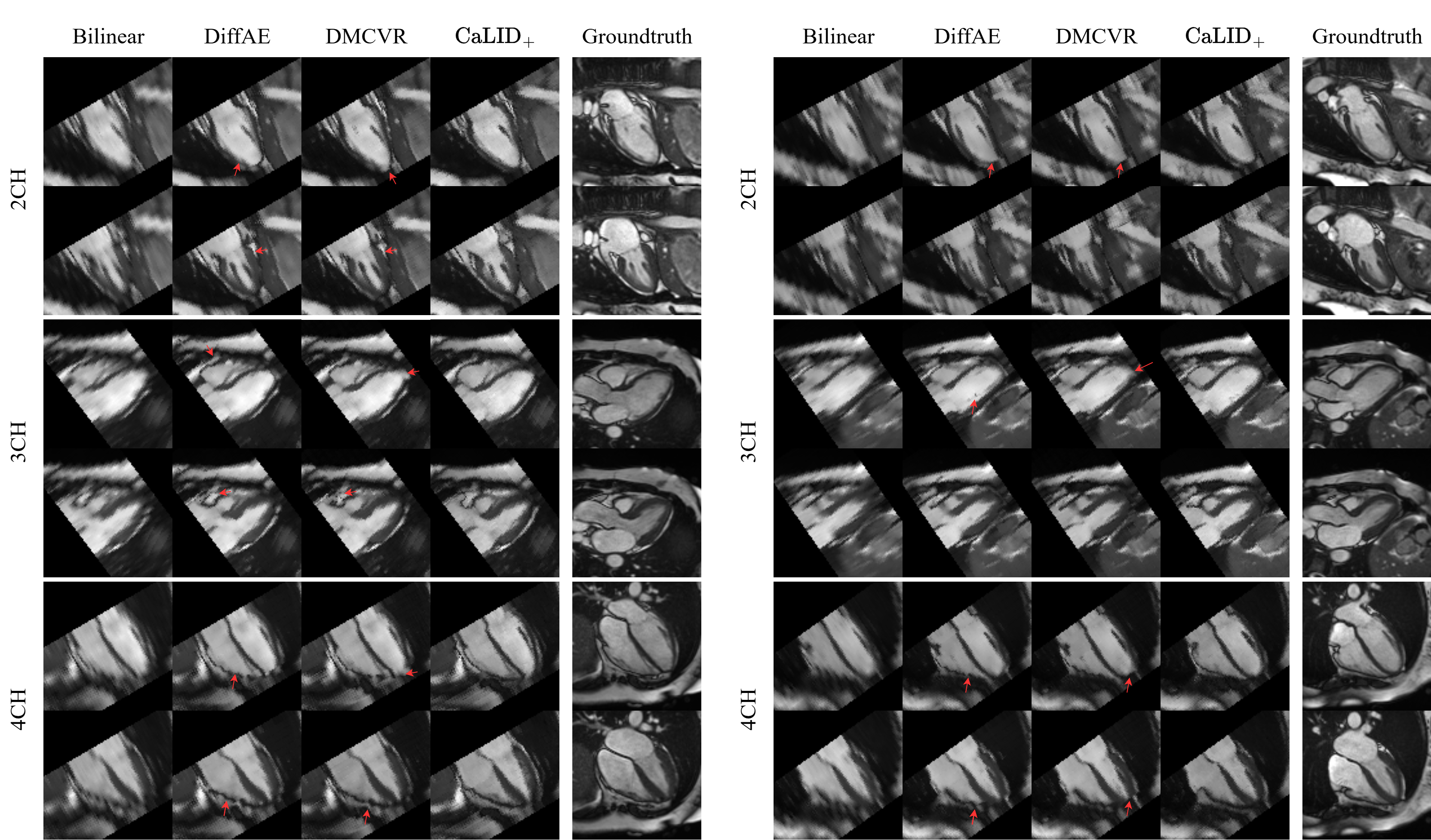}
    \caption{Visual comparison of 3D volumetric reconstructions from short-axis stacks mapped to long-axis views (2ch, 3ch, 4ch) for the end diastolic (top) and end systolic (bottom) frame. Results shown for DiffAE (128 steps), DMCVR (128 steps), and our method (8 steps). Red arrows highlight points of interest best viewed zoomed in.}
    \label{fig:sax_to_lax}
\end{figure*}

To assess 3D consistency in a physically meaningful way, we reformat the reconstructed SAX volume into corresponding long-axis (LAX) planes (Figure~\ref{fig:sax_to_lax}) and compare them against natively acquired LAX images. Because LAX is orthogonal to SAX sampling, continuity and anatomical plausibility in these reformations provide a direct proxy for inter-slice 3D coherence. We interpret this analysis qualitatively, since (i) LAX segmentation labels are not available, and (ii) slice spacing/orientation differences across acquisitions make voxel-wise and feature-wise error metrics unreliable even after geometric transformation.

In addition, we report a downstream segmentation-based analysis to probe morphological fidelity of the reconstructed volumes. We emphasize that segmentation/mesh-based cardiac modeling is not the optimization target of this work; it is used only as a complementary downstream assessment, while the primary evaluation remains in the voxel domain.

\noindent\textbf{\textit{Temporal Coherence.}}
To qualitatively assess temporal coherence, we independently apply each 2D baseline method to individual frames of a spatiotemporal (2D+T) sequence. To facilitate a meaningful comparison with our temporally adapted model, we extract temporal cross-sections along fixed spatial coordinates. Specifically, horizontal and vertical axes—across the time dimension. These cross-sectional slices reveal the temporal evolution of spatial features, enabling a direct evaluation of temporal consistency and stability across successive frames as shown in Figure~\ref{fig:temporal_consistency}.

\noindent\textbf{\textit{Ablation of the Conditioning Mechanism.}}
To evaluate the design of the conditioning mechanism, we replace the learned encoder $\tau_{\theta 1}$ with the pretrained VAE encoder $\mathcal{E}$ while keeping the subsequent mapping $\tau_{\theta 2}$ unchanged. Specifically, the VAE-based alternative is defined as
\begin{equation}
c^* = \tau_{\theta 2}\Big(\mathcal{E}\big(s^{(n-1)}\big), \mathcal{E}\big(s^{(n+1)}\big), z_t\Big)
\end{equation}
where $\mathcal{E}$ denotes the pretrained VAE encoder. This formulation isolates the impact of learning a task-specific encoding of the neighboring slices versus using a fixed latent representation from the VAE.

\begin{figure}[t]
\centering
\includegraphics[width=0.49\textwidth]{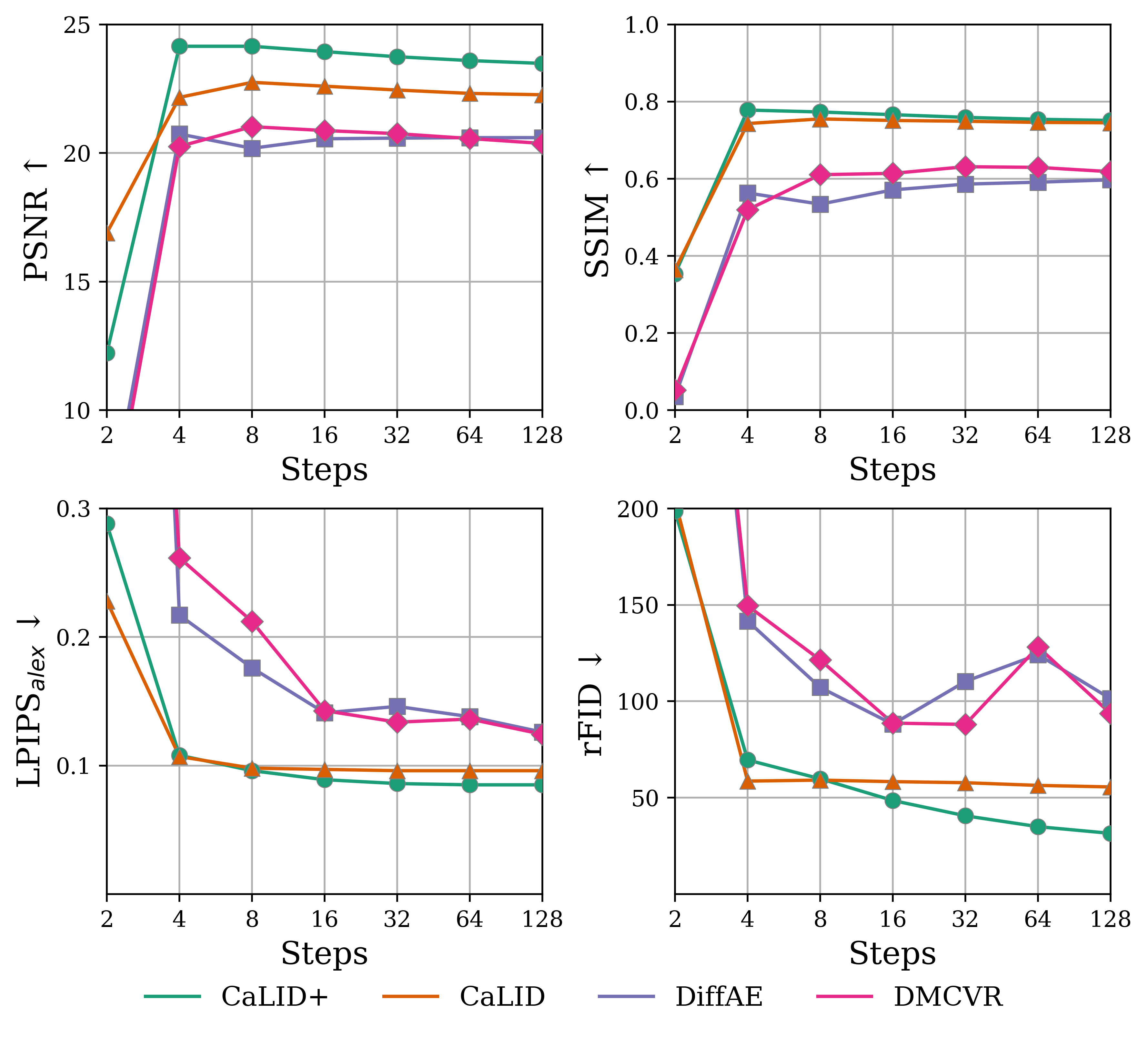}
\caption{Impact of diffusion step count on reconstruction quality, measured across PSNR, SSIM, LPIPS$_\text{alex}$, and rFID. Results illustrate how performance varies with increasing diffusion steps, highlighting trade-offs between visual performance and computational cost.}
\label{fig:metrics-vertical}
\end{figure}

\begin{figure*}[t]
    \centering
    \begin{minipage}[t]{0.49\textwidth}
        \centering
        \includegraphics[width=\textwidth]{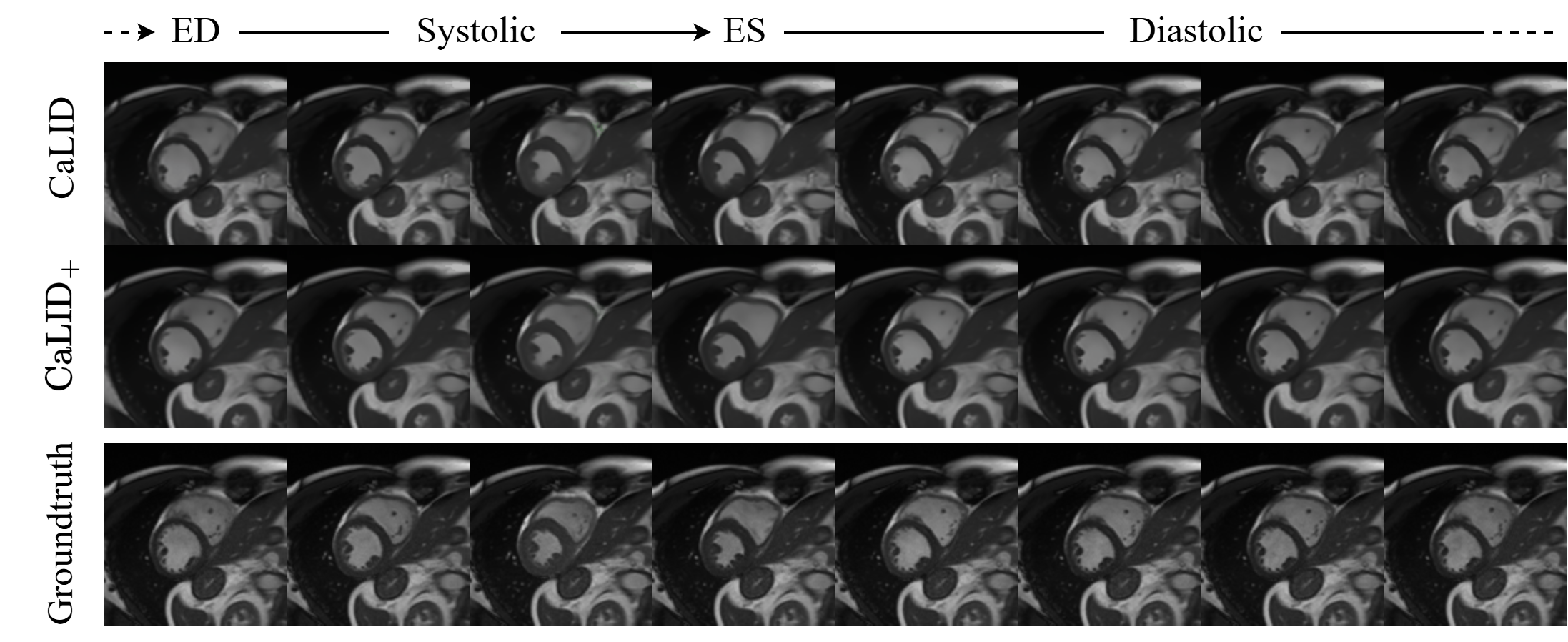}
        \caption{Comparison of 2D+T model outputs across different phases of the cardiac cycle. The temporal coherence and anatomical consistency of each method are visualized over time.}
        \label{fig:cardiac_cycle}
    \end{minipage}
    \hfill
    \begin{minipage}[t]{0.49\textwidth}
        \centering
        \includegraphics[width=\textwidth]{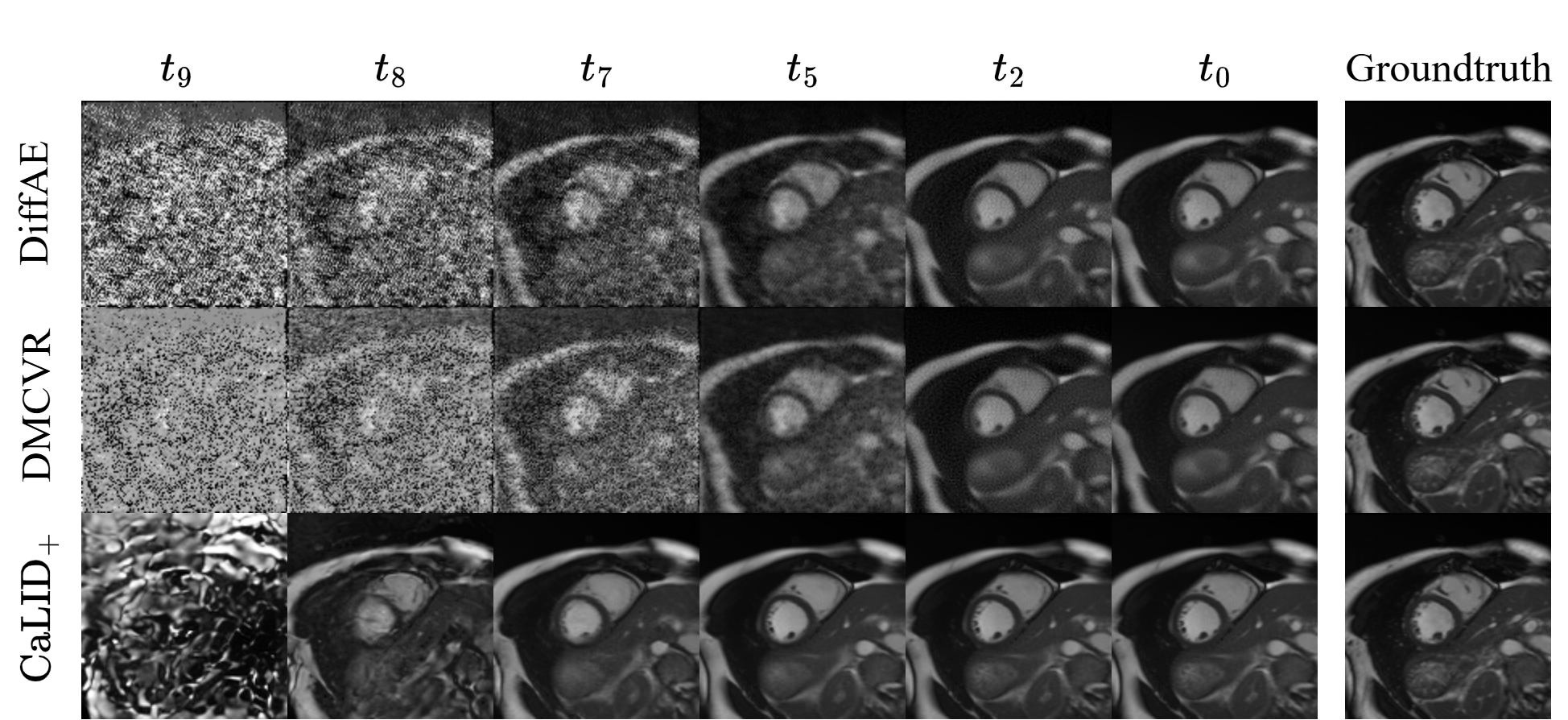}
        \caption{Denoising progression of images over a 10-step sampling process, with pure noise at the timestep $T=10$. The image shows the predicted means $\hat{x}_0^t$ for each of the timesteps $t_{9,8,7,5,2,0}$.}
        \label{fig:pred_all}
    \end{minipage}
\end{figure*}

\begin{table}[t]
\centering
\caption{Segmentation performance comparison on 2D SAX cardiac MRI. `Original' denotes results from applying a pre-trained model to unprocessed sparse slices. All metrics are evaluated against ground truth labels, using the same model for consistency.}
\begin{tabular}{c|c|c|c|c|c}
\hline 
Reg. & Metric & DiffAE & DMCVR & $\text{CaLID}_+$ & \cellcolor{lightgray}Original \\
\hline 
\multirow{4}{*}{Mean}
& Dice
  & $0.82_{\pm 0.08}$ & $0.82_{\pm 0.08}$ & $\boldsymbol{0.86_{\pm 0.09}}$ & \cellcolor{lightgray}$0.96_{\pm 0.12}$ \\
\hhline{~|-----}
& ASD
  & $1.15_{\pm 0.97}$ & $1.18_{\pm 0.99}$ & $\boldsymbol{0.85_{\pm 0.88}}$ & \cellcolor{lightgray}$0.33_{\pm 0.43}$ \\
\hhline{~|-----}
& HD
  & $3.67_{\pm 1.61}$ & $3.68_{\pm 1.69}$ & $\boldsymbol{2.82_{\pm 1.58}}$ & \cellcolor{lightgray}$1.84_{\pm 3.00}$ \\ 
\hhline{~|-----}
& ASSD
  & $1.15_{\pm 0.92}$ & $1.18_{\pm 0.97}$ & $\boldsymbol{0.90_{\pm 0.83}}$ & \cellcolor{lightgray}$0.35_{\pm 0.60}$ \\
\hline
\multirow{4}{*}{LVC}
& Dice
  & $0.90_{\pm 0.05}$ & $0.90_{\pm 0.05}$ & $\boldsymbol{0.92_{\pm 0.06}}$ & \cellcolor{lightgray}$0.98_{\pm 1.36}$ \\
\hhline{~|-----}
& ASD
  & $1.00_{\pm 1.30}$ & $1.00_{\pm 1.29}$ & $\boldsymbol{0.82_{\pm 1.23}}$ & \cellcolor{lightgray}$0.25_{\pm 0.49}$ \\
\hhline{~|-----}
& HD
  & $2.46_{\pm 1.84}$ & $2.51_{\pm 1.84}$ & $\boldsymbol{2.05_{\pm 1.81}}$ & \cellcolor{lightgray}$1.24_{\pm 3.48}$ \\
\hhline{~|-----}
& ASSD
  & $1.01_{\pm 1.22}$ & $1.02_{\pm 1.26}$ & $\boldsymbol{0.83_{\pm 1.16}}$ & \cellcolor{lightgray}$0.26_{\pm 0.73}$ \\
\hline
\multirow{4}{*}{LVM}
& Dice
  & $0.75_{\pm 0.15}$ & $0.75_{\pm 0.15}$ & $\boldsymbol{0.80_{\pm 0.16}}$ & \cellcolor{lightgray}$0.95_{\pm 0.12}$ \\
\hhline{~|-----}
& ASD
  & $0.87_{\pm 1.11}$ & $0.86_{\pm 1.47}$ & $\boldsymbol{0.72_{\pm 1.06}}$ & \cellcolor{lightgray}$0.26_{\pm 1.35}$ \\
\hhline{~|-----}
& HD
  & $3.11_{\pm 1.82}$ & $3.09_{\pm 2.13}$ & $\boldsymbol{2.51_{\pm 1.78}}$ & \cellcolor{lightgray}$1.44_{\pm 4.11}$ \\
\hhline{~|-----}
& ASSD
  & $0.89_{\pm 1.05}$ & $0.89_{\pm 1.46}$ & $\boldsymbol{0.76_{\pm 1.00}}$ & \cellcolor{lightgray}$0.27_{\pm 1.45}$ \\
\hline
\multirow{4}{*}{RVC}
& Dice
  & $0.82_{\pm 0.06}$ & $0.82_{\pm 0.05}$ & $\boldsymbol{0.87_{\pm 0.06}}$ & \cellcolor{lightgray}$0.94_{\pm 0.17}$ \\
\hhline{~|-----}
& ASD
  & $1.58_{\pm 0.70}$ & $1.68_{\pm 0.58}$ & $\boldsymbol{1.01_{\pm 0.58}}$ & \cellcolor{lightgray}$0.48_{\pm 0.81}$ \\
\hhline{~|-----}
& HD
  & $5.45_{\pm 1.95}$ & $5.46_{\pm 1.84}$ & $\boldsymbol{3.89_{\pm 2.16}}$ & \cellcolor{lightgray}$2.82_{\pm 5.83}$ \\
\hhline{~|-----}
& ASSD
  & $1.57_{\pm 0.64}$ & $1.63_{\pm 0.52}$ & $\boldsymbol{1.10_{\pm 0.54}}$ & \cellcolor{lightgray}$0.52_{\pm 1.17}$ \\
\hline
\end{tabular}
\label{table:segmentation_comp}
\end{table}

\begin{table*}[t]
\centering
\caption{Reconstruction performance under increasing interslice distance ($32\rightarrow48\,\mathrm{mm}$). All methods degrade with distance.}
\label{tab:offset_comparison}
\begin{tabular}{l|ccc|ccc}
\hline
 & \multicolumn{6}{c}{Interslice Distance} \\
Model 
& \multicolumn{3}{c}{$d=32mm$} 
& \multicolumn{3}{c}{$d=48mm$} \\
\cline{2-7}
 & PSNR $\uparrow$ & SSIM $\uparrow$ & LPIPS\textsubscript{alex} $\downarrow$
 & PSNR $\uparrow$ & SSIM $\uparrow$ & LPIPS\textsubscript{alex} $\downarrow$ \\
\hline

Bilinear Pixel 
& $15.930_{\pm 1.58}$ & $0.276_{\pm 0.05}$ & $0.222_{\pm 0.03}$
& $14.527_{\pm 1.66}$ & $0.222_{\pm 0.05}$ & $0.277_{\pm 0.03}$ \\ 

Bilinear Latent 
& $15.375_{\pm 1.65}$ & $0.351_{\pm 0.05}$ & $0.238_{\pm 0.03}$
& $13.965_{\pm 1.76}$ & $0.281_{\pm 0.05}$ & $0.295_{\pm 0.03}$ \\

DiffAE~\cite{preechakul2021diffusion}
& $17.422_{\pm 1.57}$ & $0.457_{\pm 0.05}$ & $\underline{0.183_{\pm 0.03}}$
& $\boldsymbol{15.967_{\pm 1.58}}$ & $\underline{0.388_{\pm 0.05}}$ & $\boldsymbol{0.220_{\pm 0.03}}$ \\

DMCVR~\cite{he2023dmcvr}
& $16.520_{\pm 1.56}$ & $0.437_{\pm 0.05}$ & $0.198_{\pm 0.03}$
& $14.729_{\pm 1.62}$ & $0.336_{\pm 0.05}$ & $0.258_{\pm 0.03}$ \\

CaLID  
& $\underline{18.388_{\pm 1.65}}$ & $\underline{0.535_{\pm 0.06}}$ & $\boldsymbol{0.166_{\pm 0.03}}$
& $15.427_{\pm 1.59}$ & $0.379_{\pm 0.06}$ & $\underline{0.245_{\pm 0.03}}$ \\

CaLID\textsubscript{+}  
& $\boldsymbol{19.086_{\pm 1.73}}$ & $\boldsymbol{0.576_{\pm 0.06}}$ & $0.187_{\pm 0.03}$
& $\underline{15.868_{\pm 1.64}}$ & $\boldsymbol{0.408_{\pm 0.06}}$ & $0.280_{\pm 0.03}$ \\ 
\hline
\end{tabular}
\end{table*}

\begin{table*}[t]
\centering
\caption{Reconstruction under non-uniform sampling ($\gamma=\tfrac{1}{3},\,\tfrac{1}{4}$). CaLID$^+$ achieves best performance at $\gamma=\tfrac{1}{3}$ and remains competitive at $\gamma=\tfrac{1}{4}$, indicating robustness to off-center interpolation.}
\label{tab:uniformity_comp}
\begin{tabular}{l|ccc|ccc}
\hline
 & \multicolumn{6}{c}{Uniformity} \\
Model 
& \multicolumn{3}{c}{$\gamma=\frac{1}{3}\quad , \quad d=24mm$} 
& \multicolumn{3}{c}{$\gamma=\frac{1}{4} \quad , \quad d=32mm$} \\
\cline{2-7}
 & PSNR $\uparrow$ & SSIM $\uparrow$ & LPIPS\textsubscript{alex} $\downarrow$
 & PSNR $\uparrow$ & SSIM $\uparrow$ & LPIPS\textsubscript{alex} $\downarrow$ \\
\hline

Bilinear Pixel 
& $17.735_{\pm 1.53}$ & $0.384_{\pm 0.05}$ & $0.171_{\pm 0.02}$
& $17.411_{\pm 1.60}$ & $0.381_{\pm 0.05}$ & $0.169_{\pm 0.02}$ \\ 

Bilinear Latent 
& $17.336_{\pm 1.64}$ & $0.457_{\pm 0.06}$ & $0.187_{\pm 0.02}$
& $16.951_{\pm 1.61}$ & $0.439_{\pm 0.06}$ & $0.182_{\pm 0.02}$ \\

DiffAE~\cite{preechakul2021diffusion}
& $\underline{19.143_{\pm 1.58}}$ & $\underline{0.544_{\pm 0.06}}$ & $\underline{0.151_{\pm 0.02}}$
& $\mathbf{18.830_{\pm 1.60}}$ & $\mathbf{0.537_{\pm 0.06}}$ & $\mathbf{0.158_{\pm 0.03}}$ \\

DMCVR~\cite{he2023dmcvr}   
& $18.638_{\pm 1.58}$ & $0.541_{\pm 0.06}$ & $0.154_{\pm 0.02}$
& $\underline{18.264_{\pm 1.56}}$ & $\underline{0.520_{\pm 0.06}}$ & $\underline{0.161_{\pm 0.02}}$ \\

CaLID\textsubscript{+}  
& $\mathbf{19.816_{\pm 1.71}}$ & $\mathbf{0.608_{\pm 0.06}}$ & $\mathbf{0.144_{\pm 0.03}}$
& $17.897_{\pm 1.68}$ & $0.513_{\pm 0.07}$ & $0.188_{\pm 0.03}$ \\ 

\hline
\end{tabular}
\end{table*}

% \begin{table*}[t]
% \centering
% \caption{Quantitative comparison across different offset values.
% Higher is better for PSNR and SSIM, while lower is better for LPIPS and rFID.}
% \label{tab:offset_comparison}
% \begin{tabular}{l|cccc|cccc}
% \hline
%  & \multicolumn{8}{c}{Uniformity} \\
% Model 
% & \multicolumn{4}{c}{$\alpha=\frac{1}{3}\quad , \quad d=24mm$} 
% & \multicolumn{4}{c}{$\alpha=\frac{1}{4} \quad , \quad d=32mm$} \\
% \cline{2-9}
%  & PSNR $\uparrow$ & SSIM $\uparrow$ & LPIPS $\downarrow$ & rFID $\downarrow$
%  & PSNR $\uparrow$ & SSIM $\uparrow$ & LPIPS $\downarrow$ & rFID $\downarrow$ \\
% \hline

% Bilinear Pixel 
% & $17.735_{\pm 1.53}$ & $0.384_{\pm 0.05}$ & $0.171_{\pm 0.02}$ & 64.862
% & $17.411_{\pm 1.60}$ & $0.381_{\pm 0.05}$ & $0.169_{\pm 0.02}$ & 52.542 \\ 

% Bilinear Latent 
% & $17.336_{\pm 1.64}$ & $0.457_{\pm 0.06}$ & $0.187_{\pm 0.02}$ & 70.626 
% & $16.951_{\pm 1.61}$ & $0.439_{\pm 0.06}$ & $0.182_{\pm 0.02}$ & 59.213 \\

% DiffAE  
% & $19.143_{\pm 1.58}$ & $0.544_{\pm 0.06}$ & $0.151_{\pm 0.02}$ & 108.962 
% & $18.830_{\pm 1.60}$ & $0.537_{\pm 0.06}$ & $0.158_{\pm 0.03}$ & 109.588 \\

% DMCVR   
% & $18.638_{\pm 1.58}$ & $0.541_{\pm 0.06}$ & $0.154_{\pm 0.02}$ & 100.835 
% & $18.264_{\pm 1.56}$ & $0.520_{\pm 0.06}$ & $0.161_{\pm 0.02}$ & 99.621  \\

% CaLID\textsubscript{+}  
% & $19.816_{\pm 1.71}$ & $0.608_{\pm 0.06}$ & $0.144_{\pm 0.03}$ & 52.321 
% & $17.897_{\pm 1.68}$ & $0.513_{\pm 0.07}$ & $0.188_{\pm 0.03}$ & 53.378 \\ 

% \hline
% \end{tabular}
% \end{table*}

\section{Results and Discussion}
\noindent\textbf{\textit{Quantitative 2D Generation.}} The quantitative results presented in Table \ref{table:metrics} demonstrate the superior performance of our method compared to baseline approaches. Both our base model CaLID and enhanced model CaLID$_+$ show significant improvements across all metrics for interpolation tasks. The notable gains in our base model highlight the effectiveness of learned latent interpolation over fixed schemes, capturing complex spatial and anatomical relationships between adjacent slices for more accurate and coherent reconstructions. The enhanced model CaLID$_+$, which incorporates an optimized initial noise map, further improves generation quality, particularly in pixel-level metrics such as PSNR and feature-based metrics like rFID. This improvement stems from reduced stochastic variation, yielding more consistent pixel-level reconstructions and in-distribution generation. Notably, semantic content remains largely stable, explaining the minimal gains in LPIPS and SSIM relative to the base model. Figure~\ref{fig:metrics-vertical} shows metric performance over different amounts of diffusion steps. Interestingly, for pixel-level metrics such as PSNR and SSIM on CaLID$_+$, performance decreases as the number of diffusion steps increases, whereas feature-based metrics like LPIPS and rFID improve with more diffusion steps. This reflects a fundamental tradeoff between pixel-level fidelity and perceptual quality: increasing diffusion steps enhances feature-level realism and semantic coherence, but may reduce exact pixel-wise accuracy.

\begin{figure*}[t]
    \centering
    \includegraphics[width=0.98\textwidth]{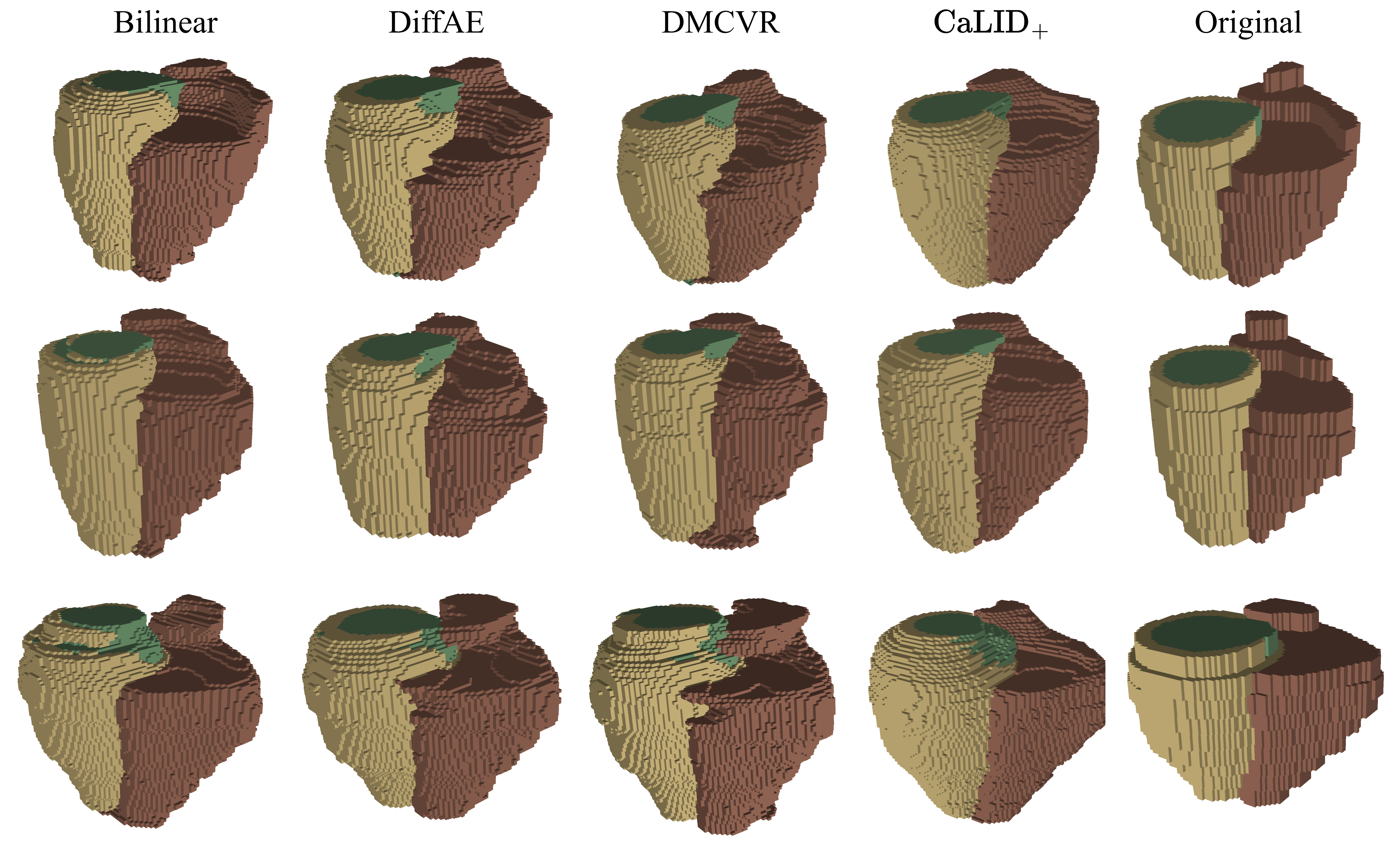}
    \caption{3D visualization of upsampled segmented whole-heart volumes obtained by applying a segmentation model~\cite{gao2022data} to the upsampled sparse volumes. Clinically relevant structures such as the LVC (green),  RVC (brown), and LVM (yellow) get visualized showcasing the improved anatomical accuracy of CaLID$_+$. Original refers to the acquired sparse stack.}
    \label{fig:3d_recon}
\end{figure*}

\noindent\textbf{\textit{Generation Quality vs. Computational Cost.}}
We evaluate the efficiency of our image generation method by analyzing the tradeoff between image quality and computational demands, a critical consideration for clinical applications. Figure~\ref{fig:metrics-vertical} illustrates the relationship between different metrics and the number of diffusion steps used to sample. Our approach already achieves convergence after only four steps and is significantly outperforming previous methods that require substantially more steps for comparable quality. Inference time scales linearly with the number of denoising steps, substantially reducing compute time.

CaLID$_+$ starts closer to the data manifold, so later steps refine structure rather than pixels: perceptual metrics (LPIPS, rFID) improve while pixel metrics (PSNR, SSIM) decrease. Standard CaLID, initialized from Gaussian noise, mainly adjusts textures and shows stable perceptual gains.

Additionally, Figure~\ref{fig:pred_all} illustrates the denoising progression of images over a 10-step sampling process, with pure noise corresponding to timestep $T=10$. Each image represents the predicted mean $\hat{x}_0^t$ at the respective timestep $t$, highlighting the progressive reduction of noise and facilitating a comparison of residual noise levels throughout the sampling trajectory.

The results highlight our method's superior performance, achieving high-detailed outputs after just four steps, and thus enabling faster denoising while maintaining higher fidelity. Furthermore, given that our base model CaLID can interpolate from random starting noise, there is no mandatory need for diffusion inversion interpolation during inference, leading to a 3 $\times$ accelerated slice generation by design as also noted in Table~\ref{table:metrics}. 

\noindent\textbf{\textit{Quantitative 2D Morphology Assessment.}} We evaluate the morphological fidelity of our interpolated cardiac images using a downstream segmentation task as a proxy measure. Table~\ref{table:segmentation_comp} summarizes the segmentation results obtained by MedFormer, where the “original” column reflects the segmentation model’s performance on unprocessed images of the sparse acquisition, establishing an upper bound as a reference. Because MedFormer is trained exclusively on real CMR data, segmentation performance drops if the generated images deviate significantly from the true data distribution. Consequently, a score closer to the “original” column indicates more realistic synthetic slices. Our method consistently outperforms the baselines across all cardiac regions on every metric, demonstrating that the images generated by $\text{CaLID}_+$ align more closely with real CMR data. This morphological fidelity is essential for clinical tasks that require accurate structural information, highlighting the practical diagnostic potential of our approach.

\begin{table*}[h]
\centering
\caption{MAE (mean ± SD) for ventricular phenotype prediction. $\text{CaLID}_{+}^{2D}$ and $\text{CaLID}_{+}^{2D+T}$ closely match Original data as the upper bound, while DiffAE and DMCVR show higher errors.}
\begin{tabular}{c|c|c|c|c|c|c}
\hline
Method & RVESV (mL) & LVESV (mL) & RVEDV (mL) & LVEDV (mL) & RVEF (\%) & LVEF (\%) \\
\hline

DiffAE~\cite{preechakul2021diffusion}
& $\underline{6.902_{\pm 5.83}}$ & $\underline{5.821_{\pm 4.89}}$ & $\underline{9.651_{\pm 7.96}}$ & $7.105_{\pm 6.13}$ & $-$ & $-$ \\

DMCVR~\cite{he2023dmcvr}
& $7.045_{\pm 5.67}$ & $6.023_{\pm 4.96}$ & $9.941_{\pm 7.95}$ & $\underline{7.025_{\pm 6.05}}$ & $-$ & $-$ \\

$\text{CaLID}_{+}^{2D}$
& $\mathbf{6.532_{\pm 5.49}}$ & $\mathbf{5.724_{\pm 4.72}}$ & $\mathbf{9.208_{\pm 7.59}}$ & $\mathbf{6.645_{\pm 5.82}}$ & $-$ & $-$ \\

\cellcolor{lightgray}Original
& \cellcolor{lightgray}$6.552_{\pm 5.57}$
& \cellcolor{lightgray}$5.758_{\pm 4.91}$
& \cellcolor{lightgray}$8.881_{\pm 7.56}$
& \cellcolor{lightgray}$6.988_{\pm 6.07}$
& \cellcolor{lightgray}$-$ 
& \cellcolor{lightgray}$-$ \\

\hline

$\text{CaLID}_{+}^{2D+T}$
& $5.424_{\pm4.58}$ & $4.774_{\pm 3.97}$ & $7.679_{\pm 3.97}$ & $7.375_{\pm 5.89}$ & $3.331_{\pm 2.69}$ & $3.062_{\pm 2.42}$ \\

\cellcolor{lightgray}Original
& \cellcolor{lightgray}$5.177_{\pm 4.33}$
& \cellcolor{lightgray}$4.569_{\pm 3.79}$
& \cellcolor{lightgray}$7.572_{\pm 6.38}$
& \cellcolor{lightgray}$7.837_{\pm 6.08}$
& \cellcolor{lightgray}$3.266_{\pm 2.55}$
& \cellcolor{lightgray}$2.996_{\pm 2.36}$ \\

\hline
\end{tabular}
\label{table:phenotype_comp_transposed}
\end{table*}

\begin{figure*}[t]
    \centering
    \includegraphics[width=0.98\textwidth]{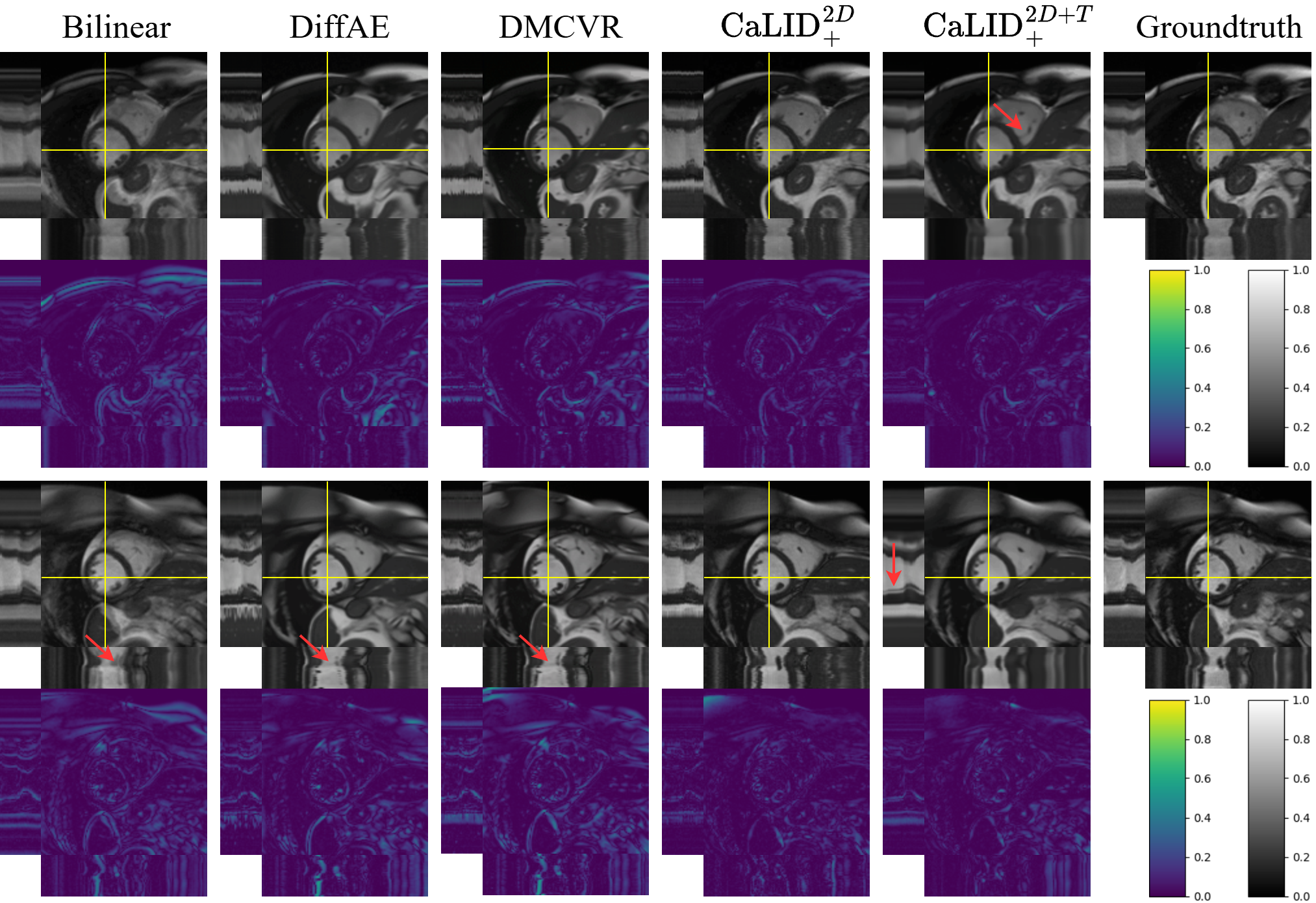}
    \caption{Temporal coherence of different methods applied to a 2D+T sequence. Yellow axes indicate the cross-sectional planes used to visualize temporal consistency: the bottom row of the slice shows the temporal evolution along the horizontal axis, while the left column shows the evolution along the vertical axis. Error maps are displayed in veridian absolute color mapping and failure cases are highlighted using red arrows.}
    \label{fig:temporal_consistency}
\end{figure*}

\noindent\textbf{\textit{Robustness Against Inter-Slice Gaps and Uniformity.}}
Tables~\ref{tab:offset_comparison} and~\ref{tab:uniformity_comp} evaluate reconstruction across increasing interslice gaps and off-center interpolation. Performance degrades with larger distances ($32\rightarrow48\,\mathrm{mm}$) for all methods; however, CaLID variants consistently outperform classical interpolation and remain competitive with diffusion baselines. At $32\,\mathrm{mm}$, CaLID$_+$ achieves the highest PSNR and SSIM while CaLID attains the lowest LPIPS. At $48\,\mathrm{mm}$ diffusion models slightly improve pixel-wise fidelity but CaLID preserves lower perceptual error, indicating improved structural consistency. Uniformity experiments further show that CaLID$_+$ performs best at $\gamma=\tfrac{1}{3}$ and remains competitive at $\gamma=\tfrac{1}{4}$.
We note that, as with all generative approaches, hallucination becomes an increasing concern as interslice distances grow, particularly when extrapolating beyond the training distribution. Moreover, non-uniform slice acquisition and large interslice gaps are not standard clinical practice. The model is trained using neighboring slices at $d=16\,\mathrm{mm}$ due to limited data availability, corresponding to a practical upsampling use case of $8\,\mathrm{mm}$. Consequently, the larger-distance and non-uniform settings intentionally deviate from the target deployment scenario and should be interpreted primarily as stress tests of robustness rather than representative clinical configurations.

\noindent\textbf{\textit{Downstream Phenotype Prediction.}}
Across all ventricular volume targets, the generative models produced prediction errors comparable to the real (Original) data distribution. In the 2D setting, $\text{CaLID}_{+}^{2D}$ consistently achieved the lowest errors across all volumetric phenotypes and closely matched the performance obtained on Original data, while DiffAE and DMCVR showed slightly higher errors, indicating weaker preservation of phenotype-relevant information (see Table~\ref{table:phenotype_comp_transposed}). Notably, errors on CaLID samples were in some cases even lower than on real data, which does not imply improved biological accuracy but rather reflects a denoising and manifold-regularization effect: the generator suppresses acquisition noise and anatomically inconsistent variations, making the phenotype more deterministically encoded and therefore easier to predict.
In the temporally-aware setting (2D+T), baseline prediction errors on real data remained low and additionally enabled accurate estimation of functional indices, establishing a strong reference for temporal phenotype prediction. Overall, using CaLID, enables the reduction of acquisition slices while still maintaining competitive phenotype prediction.

\noindent\textbf{\textit{Qualitative 3D Reconstruction Assessment.}}\label{chapter:3dres} This section presents a qualitative visual evaluation of our whole-heart volume reconstruction methodology. Figure~\ref{fig:sax_to_lax} provides a comparative analysis between our supersampled short-axis stack, mapped to corresponding long-axis views (2ch, 3ch, 4ch), and the ground truth images. Our method demonstrates detail-preserved cardiac structures and sharp edge definition along the long-axis plane. Better reconstruction quality can especially be seen for the highlighted regions including the papillary muscles and the inferior cardiac wall. Notably, our approach achieves these results using only 8 diffusion steps while DiffAE requires 64 steps and DMCVR 128 steps. This highlights a substantial improvement in computational efficiency, with $\text{CaLID}$ achieving up to a 24x and $\text{CaLID}_+$ a 8x speedup in 3D generation compared to prior approaches. Notably, models such as DiffAE and DMCVR exhibit failure cases when the number of diffusion steps is reduced, as their performance deteriorates under accelerated sampling. We hypothesize that this is due to error accumulation during autoregressive interpolation: with each bisectional generation, small deviations compound, and eventually the intermediate representations drift away from the data manifold learned by the diffusion model. In contrast, our method demonstrates greater robustness to such compounding errors given our dense manifold and strict conditioning assumption, as evidenced by its consistently lower rFID scores. This suggests that our model maintains better alignment with the target distribution throughout the autoregressive process, enabling high-quality synthesis even under reduced inference budgets. 

\noindent Figure~\ref{fig:3d_recon} presents a comparative analysis of three-dimensional reconstructions derived from segmented cardiac slices, illustrating the strengths of our proposed method relative to existing interpolation approaches. Traditional techniques such as bilinear interpolation fail to capture high-frequency anatomical details, often resulting in aliasing and staircase artifacts along structural boundaries. While generative models like DiffAE and DMCVR offer more plausible reconstructions, they still exhibit limitations in preserving fine-grained features, particularly in regions requiring precise delineation of cardiac anatomy. In contrast, our method demonstrates superior spatial fidelity, maintaining both sharp anatomical contours and coherent inter-slice transitions. This leads to volumetric reconstructions that more faithfully reflect the true morphology of cardiac structures and support clinically relevant interpretation.

\noindent\textbf{\textit{Spatiotemporal Flexibility}}
Our findings demonstrate that the proposed method generalizes effectively to temporally-resolved 2D+T cardiac MRI sequences, highlighting the flexibility and scalability of the latent diffusion framework for spatiotemporal interpolation tasks. Quantitative evaluations show that both CaLID and CaLID$_+$ achieve strong performance across standard metrics, reflecting accurate reconstruction of intermediate frames and robust temporal consistency. Qualitative analysis further confirms that the interpolated frames maintain close adherence to anatomical structures throughout the cardiac cycle (see Figure~\ref{fig:cardiac_cycle}), including key phases such as end-diastole (ED) and end-systole (ES).

Notably, CaLID$^{2D}_+$ exhibits superior preservation of fine-grained anatomical features, such as the papillary muscles, which are particularly challenging due to their intricate morphology and subtle temporal dynamics. Figure~\ref{fig:temporal_consistency} reinforces this finding. In the horizontal temporal cross-section, only our models reconstruct the papillary muscle correctly across timeframes; the baselines fail to preserve it. This improvement comes with a trade-off, however. CaLID$_+^{2D+T}$ produces the smoothest and most temporally consistent reconstructions, but it is less accurate on fine spatial detail: it tends to visually merge spatially close structures, blending neighboring anatomies such as papillary muscles or trabeculations. Along the temporal axis, these same structures also tend to appear elongated. Such localized failures mark the point at which the added temporal fidelity trades against the exact reconstruction of fine cardiac detail. Overall, though, the improved representation of dynamic anatomical structure yields more realistic and clinically meaningful reconstructions, capturing the complex cardiac motion present in time-resolved MRI data.

\begin{figure*}[t]
    \centering
    \includegraphics[width=0.98\textwidth]{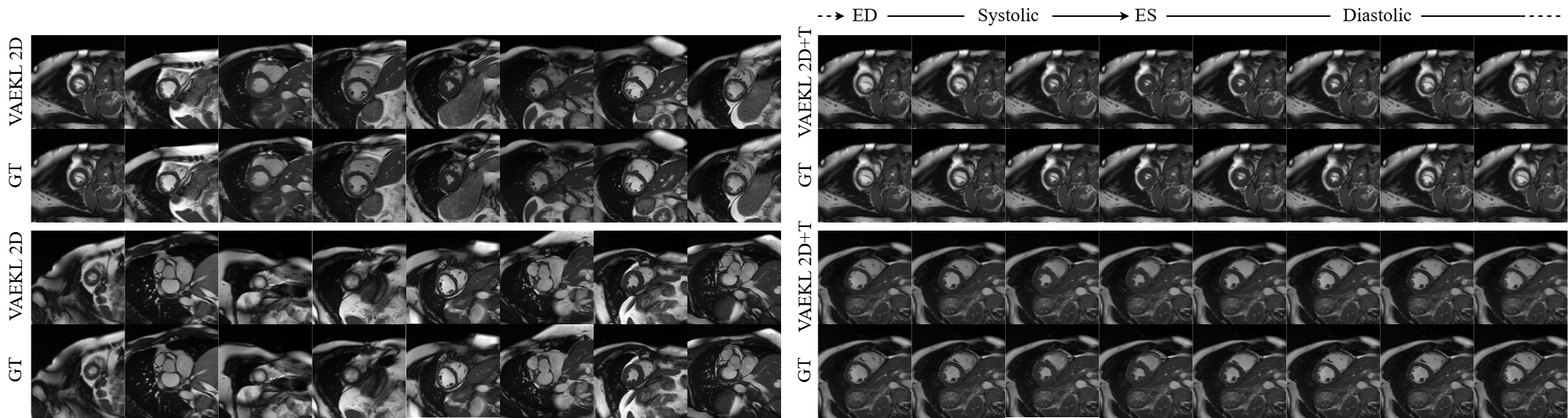}    \caption{VAE reconstructions: VAEKL 2D+T (right) produces smoother, temporally consistent outputs, while VAEKL 2D (left) preserves sharper high-frequency details.}
    \label{fig:autoencoders}
\end{figure*}

\begin{table*}[t]
\centering

\begin{minipage}{0.48\textwidth}
\centering
\caption{Reconstruction metrics: VAEKL 2D+T improves PSNR/SSIM but worsens LPIPS and rFID, indicating smoother yet less sharp outputs than VAEKL 2D.}
\renewcommand{\arraystretch}{1.1}
\resizebox{\textwidth}{!}{%
\begin{tabular}{l|cccc}
\hline
\textbf{Model} & \textbf{PSNR} $\uparrow$ & \textbf{SSIM} $\uparrow$ & \textbf{LPIPS\textsubscript{alex}} $\downarrow$ & \textbf{rFID} $\downarrow$ \\
\hline
VAEKL 2D   & $33.473_{\pm 1.108}$ & $0.938_{\pm 0.006}$ & $0.017_{\pm 0.003}$ & $17.129$ \\
VAEKL 2D+T & $35.596_{\pm 1.031}$ & $0.9640_{\pm 0.007}$ & $0.031_{\pm 0.008}$ & $46.725$ \\
\hline
\end{tabular}
}
\label{tab:autoencoder_comparison}
\end{minipage}
\hfill
\begin{minipage}{0.48\textwidth}
\centering
\caption{Comparison of conditioning strategies on reconstruction and perceptual metrics.}
\renewcommand{\arraystretch}{1.1}
\resizebox{\textwidth}{!}{%
\begin{tabular}{lccccc}
\hline
\textbf{Model} & \textbf{Cond.} & \textbf{PSNR} $\uparrow$ & \textbf{SSIM} $\uparrow$ & \textbf{LPIPS\textsubscript{alex}} $\downarrow$ & \textbf{rFID} $\downarrow$ \\
\hline
CaLID   & VAE & $\boldsymbol{23.157_{\pm 2.31}}$ & $0.741_{\pm 0.07}$ & $0.104_{\pm 0.03}$ & $66.370$ \\
CaLID   & Ours  & ${22.748_{\pm 2.14}}$ & $\boldsymbol{0.755_{\pm 0.07}}$ & $\boldsymbol{0.097_{\pm 0.02}}$ & $\boldsymbol{55.585}$ \\
\hline
CaLID\textsubscript{+} & VAE & $23.986_{\pm 2.40}$ & $0.776_{\pm 0.07}$ & $0.106_{\pm 0.03}$ & $39.894$ \\
CaLID\textsubscript{+} & Ours & $\boldsymbol{24.160_{\pm 2.35}}$ & $\boldsymbol{0.778_{\pm 0.07}}$ & $\boldsymbol{0.086_{\pm 0.02}}$ & $\boldsymbol{31.508}$ \\
\hline
\end{tabular}
}
\label{tab:conditioning_comparison}
\end{minipage}

\end{table*}

\noindent\textbf{\textit{Variational Autoencoders.}}
We evaluate the pretrained autoencoder to characterize its latent representation. The temporally-aware variant (VAEKL 2D+T) produces smoother reconstructions with greater temporal consistency. This smoothness is most apparent in large, uniform regions, where high frequencies are suppressed, but it comes at the cost of perceptual sharpness, reflected in higher LPIPS and rFID (see Table~\ref{tab:autoencoder_comparison}). In contrast, the spatial model (VAEKL 2D) preserves higher-frequency structures. Figure~\ref{fig:autoencoders} illustrates this trade-off, while both models maintain anatomically coherent latent representations suitable for diffusion-based interpolation.

\noindent\textbf{\textit{Ablation of the Conditioning Mechanism.}}
Replacing the VAE conditioning with our approach consistently improves perceptual fidelity and distribution alignment as shown in Table~\ref{tab:conditioning_comparison}. Across both CaLID variants, our conditioning yields higher SSIM and lower LPIPS, indicating better structural and perceptual quality. Most notably, rFID is substantially reduced, showing that generated samples lie closer to the ground-truth data manifold. While the base CaLID model exhibits a slight PSNR trade-off, the enhanced CaLID\textsubscript{+} benefits across all metrics, confirming improved reconstruction and realism.

\noindent\textbf{\textit{Limitations.}}
While our method demonstrates strong performance, it is inherently subject to the common tradeoff observed in generative models between pixel-level accuracy and perceptual fidelity. Achieving high perceptual quality and anatomically plausible interpolations can sometimes come at the expense of exact pixel-wise reconstruction fidelity. While potential remedies could include incorporating stronger anatomical priors, this would yet again introduce dependence on auxiliaries intentionally avoided in this work to simplify the pipeline and clinical application. Furthermore, the quality and consistency of input slices play a crucial role in the model's output; severe motion artifacts, inconsistent slice spacing, or pronounced inter-slice anatomical variations can degrade reconstruction quality and spatiotemporal coherence. The model was trained exclusively on ventricular short-axis (SAX) images, limiting consistent reconstruction of full atrial anatomy. Additionally, the current dataset does not contain paired standard 3D cine CMR acquisitions, preventing direct qualitative and quantitative comparison against fully acquired volumetric ground truth and limiting the ability to assess absolute volumetric accuracy relative to native 3D imaging. Such challenges remain critical considerations when applying the model to clinical data, where image acquisition variability is unavoidable.

\noindent\textbf{\textit{Outlook.}}
Although our framework is currently validated on cardiac MRI sequences, its underlying architecture and latent diffusion approach are broadly applicable. We anticipate that other imaging modalities characterized by limited resolution or sparse sampling such as low-resolution brain MRI, computed tomography (CT), or dynamic ultrasound could similarly benefit from this approach. Future work will focus on adapting and evaluating the method across diverse modalities to enhance its clinical impact and generalizability. Furthermore, the incorporation of long-axis views could potentially benefit the anatomical plausibility and will be studied as future work. 

\section{Conclusion}
\noindent In conclusion, this study introduces a novel framework for whole-heart volume reconstruction based on a latent diffusion model trained specifically for interpolation. The proposed method achieves superior performance across both quantitative and qualitative metrics, outperforming existing state-of-the-art approaches in terms of image fidelity, anatomical precision, volumetric consistency and robustness. Furthermore, the model's computational efficiency and minimal input requirements highlight its practical applicability in clinical and research settings. These strengths position the approach as a robust solution for high-quality reconstruction of 3D and 3D+T cardiac volumes, while also establishing a solid foundation for future developments in downstream tasks such as functional assessment, motion analysis, and disease characterization.


\begin{thebibliography}{00}

\bibitem{bai2020population}
W. Bai~\emph{et al.},
``A population-based phenome-wide association study of cardiac and aortic structure and function,''
\textit{Nature Medicine}, vol. 26, no. 10, pp. 1654--1662, 2020.

\bibitem{wang2024screening}
Y.-R. Wang~\emph{et al.},
``Screening and diagnosis of cardiovascular disease using artificial intelligence-enabled cardiac magnetic resonance imaging,''
\textit{Nature Medicine}, vol. 30, no. 5, pp. 1471--1480, 2024.

\bibitem{zhang2025towards}
Y. Zhang~\emph{et al.},
``Towards cardiac mri foundation models: Comprehensive visual-tabular representations for whole-heart assessment and beyond,''
\textit{arXiv preprint arXiv:2504.13037}, 2025.

\bibitem{zhang2024whole}
Y. Zhang, C. Chen, S. Shit, S. Starck, D. Rueckert, and J. Pan,
``Whole heart 3d+ t representation learning through sparse 2d cardiac mr images,''
in \textit{International Conference on Medical Image Computing and Computer-Assisted Intervention}, 
pp. 359--369, 2024.

\bibitem{chang2022deeprecon}
Q. Chang~\emph{et al.},
``Deeprecon: Joint 2d cardiac segmentation and 3d volume reconstruction via a structure-specific generative method,''
in \textit{International Conference on Medical Image Computing and Computer-Assisted Intervention}, 
pp. 567--577, 2022.

\bibitem{jayakumar2023sadir}
N.~Jayakumar, T.~Hossain, and M.~Zhang,
``SADIR: Shape-Aware Diffusion Models for 3D Image Reconstruction,''
in \emph{International Workshop on Shape in Medical Imaging}. Springer, 2023, pp.~287--300.


\bibitem{ye2023neural}
M.~Ye, D.~Yang, M.~Kanski, L.~Axel, and D.~Metaxas,
``Neural Deformable Models for 3D Bi-Ventricular Heart Shape Reconstruction and Modeling from 2D Sparse Cardiac Magnetic Resonance Imaging,''
in \emph{Proceedings of the IEEE/CVF International Conference on Computer Vision (ICCV)}, 2023, pp.~14247--14256.

\bibitem{qiao2025personalized}
M.~Qiao, K.~A.~McGurk, S.~Wang, P.~M.~Matthews, D.~P.~O’Regan, and W.~Bai,
``A Personalized Time-Resolved 3D Mesh Generative Model for Unveiling Normal Heart Dynamics,''
\emph{Nature Machine Intelligence}, vol.~7, no.~5, pp.~800--811, 2025.

\bibitem{pan2024reconstruction}
J. Pan, W. Huang, D. Rueckert, T. Küstner, and K. Hammernik,
``Reconstruction-driven motion estimation for motion-compensated MR CINE imaging,''
\textit{IEEE Transactions on Medical Imaging}, 2024.

\bibitem{zhao2019motion}
N. Zhao, D. O’Connor, A. Basarab, D. Ruan, and K. Sheng,
``Motion compensated dynamic MRI reconstruction with local affine optical flow estimation,''
\textit{IEEE Transactions on Biomedical Engineering}, vol. 66, no. 11, pp. 3050--3059, 2019.

\bibitem{mun2013motion}
S. Mun and J. E. Fowler,
``Motion-compensated compressed-sensing reconstruction for dynamic MRI,''
in \textit{2013 IEEE International Conference on Image Processing}, 
pp. 1006--1010, 2013.

\bibitem{pan2024unrolled}
J. Pan, M. Hamdi, W. Huang, K. Hammernik, T. Kuestner, and D. Rueckert,
``Unrolled and rapid motion-compensated reconstruction for cardiac CINE MRI,''
\textit{Medical Image Analysis}, vol. 91, p. 103017, 2024.

\bibitem{monney2015single}
P. Monney, D. Piccini, T. Rutz, G. Vincenti, S. Coppo, S. C. Koestner, N. Sekarski, S. Di Bernardo, J. Bouchardy, M. Stuber, et al.,
\textit{Single centre experience of the application of self navigated 3D whole heart cardiovascular magnetic resonance for the assessment of cardiac anatomy in congenital heart disease},
Journal of Cardiovascular Magnetic Resonance, vol. 17, no. 1, p. 55, 2015.

\bibitem{piccini2012respiratory}
D. Piccini, A. Littmann, S. Nielles-Vallespin, M. O. Zenge,
\textit{Respiratory self-navigation for whole-heart bright-blood coronary MRI: methods for robust isolation and automatic segmentation of the blood pool},
Magnetic Resonance in Medicine, vol. 68, no. 2, pp. 571--579, 2012.

\bibitem{ogier2025free}
A. C. Ogier, I. Montón Quesada, X. Sieber, P. Calarnou, J.-B. Ledoux, B. Milani, P. Antiochos, J. Schwitter, C. W. Roy, J. Yerly, et al.,
\textit{Free-running 5D whole-heart MRI for isotropic cardiac function measurements at 3T without contrast agents},
Magnetic Resonance in Medicine, vol. 93, no. 6, pp. 2386--2400, 2025.

\bibitem{moghari2018free}
M. H. Moghari, A. Barthur, M. E. Amaral, T. Geva, A. J. Powell,
\textit{Free-breathing whole-heart 3D cine magnetic resonance imaging with prospective respiratory motion compensation},
Magnetic Resonance in Medicine, vol. 80, no. 1, pp. 181--189, 2018.

\bibitem{holtackers2025low}
R. J. Holtackers, A. C. Ogier, L. Romanin, E. Tenisch, I. M. Quesada, R. B. van Heeswijk, C. W. Roy, J. Yerly, M. Prsa, M. Stuber,
\textit{How low can we go? The effect of acquisition duration on cardiac volume and function measurements in free-running cardiac and respiratory motion-resolved five-dimensional whole-heart cine magnetic resonance imaging at 1.5 T},
Journal of Cardiovascular Magnetic Resonance, vol. 27, no. 1, p. 101863, 2025.

\bibitem{preechakul2021diffusion}
K. Preechakul, N. Chatthee, S. Wizadwongsa, and S. Suwajanakorn,
``Diffusion Autoencoders: Toward a Meaningful and Decodable Representation,''
in \textit{IEEE Conference on Computer Vision and Pattern Recognition (CVPR)}, 2022.

\bibitem{karras2020analyzing}
T. Karras, S. Laine, M. Aittala, J. Hellsten, J. Lehtinen, and T. Aila,
``Analyzing and improving the image quality of stylegan,''
in \textit{Proceedings of the IEEE/CVF Conference on Computer Vision and Pattern Recognition}, 
pp. 8110--8119, 2020.

\bibitem{stylegan}
T. Karras, S. Laine, and T. Aila,
``A Style-Based Generator Architecture for Generative Adversarial Networks,''
\textit{IEEE Transactions on Pattern Analysis and Machine Intelligence}, 
vol. 43, no. 12, pp. 4217--4228, 2021.

\bibitem{he2023dmcvr}
X. He~\emph{et al.},
``DMCVR: Morphology-Guided Diffusion Model for 3D Cardiac Volume Reconstruction,''
in \textit{International Conference on Medical Image Computing and Computer-Assisted Intervention}, 
pp. 132--142, 2023.

\bibitem{zheng2024noisediffusion}
P. Zheng, Y. Zhang, Z. Fang, T. Liu, D. Lian, and B. Han,
``NoiseDiffusion: Correcting Noise for Image Interpolation with Diffusion Models beyond Spherical Linear Interpolation,''
in \textit{The Twelfth International Conference on Learning Representations}, 2024.

\bibitem{stolt2023nisf}
N. Stolt-Ansó, J. McGinnis, J. Pan, K. Hammernik, and D. Rueckert,
``Nisf: Neural implicit segmentation functions,''
in \textit{International Conference on Medical Image Computing and Computer-Assisted Intervention}, 
pp. 734--744, 2023.

\bibitem{wang2021joint}
S. Wang~\emph{et al.},
``Joint motion correction and super resolution for cardiac segmentation via latent optimisation,''
in \textit{Medical Image Computing and Computer Assisted Intervention--MICCAI 2021: 24th International Conference, Strasbourg, France, September 27--October 1, 2021, Proceedings, Part III 24}, 
pp. 14--24, 2021.

\bibitem{leng2013medical}
J. Leng, G. Xu, and Y. Zhang,
``Medical image interpolation based on multi-resolution registration,''
\textit{Computers \& Mathematics with Applications}, vol. 66, no. 1, pp. 1--18, 2013.

\bibitem{meng2022mulvimotion}
Q. Meng~\emph{et al.},
``MulViMotion: Shape-aware 3D myocardial motion tracking from multi-view cardiac MRI,''
\textit{IEEE Transactions on Medical Imaging}, vol. 41, no. 8, pp. 1961--1974, 2022.

\bibitem{frakes2008new}
D. H. Frakes~\emph{et al.},
``A new method for registration-based medical image interpolation,''
\textit{IEEE Transactions on Medical Imaging}, vol. 27, no. 3, pp. 370--377, 2008.

\bibitem{bubeck2025reconstruct}
N. Bubeck, Y. Zhang, S. Shit, D. Rueckert, and J. Pan,
``Reconstruct or Generate: Exploring the Spectrum of Generative Modeling for Cardiac MRI,''
\textit{arXiv preprint arXiv:2507.19186}, 2025.

\bibitem{biller2026tumorflow}
V.~Biller, N.~Bubeck, L.~Zimmer, A.~C.~Erdur, S.~Nagar, A.~Meyer-Baese, D.~R{\"u}ckert, B.~Wiestler, and J.~Weidner,  
``TumorFlow: Physics-Guided Longitudinal MRI Synthesis of Glioblastoma Growth,''  
\emph{arXiv preprint arXiv:2603.04058}, 2026.
%%% 26
\bibitem{bau2019seeing}
D. Bau~\emph{et al.},
``Seeing what a gan cannot generate,''
in \textit{Proceedings of the IEEE/CVF International Conference on Computer Vision}, 
pp. 4502--4511, 2019.

\bibitem{kingma2013auto}
D. P. Kingma~\emph{et al.},
``Auto-encoding variational bayes,''
Banff, Canada, 2013.

\bibitem{esser2021taming}
P. Esser, R. Rombach, and B. Ommer,
``Taming transformers for high-resolution image synthesis,''
in \textit{Proceedings of the IEEE/CVF Conference on Computer Vision and Pattern Recognition}, 
pp. 12873--12883, 2021.


\bibitem{rezende2015variational}
D. Rezende and S. Mohamed,
``Variational inference with normalizing flows,''
in \textit{International Conference on Machine Learning}, 
pp. 1530--1538, 2015.
%% 30

\bibitem{ho2022classifier}
J. Ho and T. Salimans,
``Classifier-free diffusion guidance,''
\textit{arXiv preprint arXiv:2207.12598}, 2022.

\bibitem{dhariwal2021diffusion}
P. Dhariwal and A. Nichol,
``Diffusion models beat gans on image synthesis,''
\textit{Advances in Neural Information Processing Systems}, vol. 34, pp. 8780--8794, 2021.
%% 32
\bibitem{song2020score}
Y. Song, J. Sohl-Dickstein, D. P. Kingma, A. Kumar, S. Ermon, and B. Poole,
``Score-based generative modeling through stochastic differential equations,''
\textit{arXiv preprint arXiv:2011.13456}, 2020.

\bibitem{ho2020denoising}
J. Ho, A. Jain, and P. Abbeel,
``Denoising diffusion probabilistic models,''
\textit{Advances in Neural Information Processing Systems}, vol. 33, pp. 6840--6851, 2020.
%% 34
\bibitem{song2020denoising}
J. Song, C. Meng, and S. Ermon,
``Denoising diffusion implicit models,''
\textit{arXiv preprint arXiv:2010.02502}, 2020.
%% 35
\bibitem{rombach2022high}
R. Rombach, A. Blattmann, D. Lorenz, P. Esser, and B. Ommer,
``High-resolution image synthesis with latent diffusion models,''
in \textit{Proceedings of the IEEE/CVF Conference on Computer Vision and Pattern Recognition}, 
pp. 10684--10695, 2022.
%% 36

\bibitem{lin2024diffbir}
X. Lin~\emph{et al.},
``Diffbir: Toward blind image restoration with generative diffusion prior,''
in \textit{European Conference on Computer Vision}, 
pp. 430--448, 2024.

\bibitem{zhang2023adding}
L. Zhang, A. Rao, and M. Agrawala,
``Adding conditional control to text-to-image diffusion models,''
in \textit{Proceedings of the IEEE/CVF International Conference on Computer Vision}, 
pp. 3836--3847, 2023.
%% 38

\bibitem{ning2023elucidating}
M.~Ning, M.~Li, J.~Su, A.~A.~Salah, and I.~O.~Ertugrul,
``Elucidating the Exposure Bias in Diffusion Models,''
\emph{arXiv preprint arXiv:2308.15321}, 2023.

\bibitem{ning2023input}
M.~Ning, E.~Sangineto, A.~Porrello, S.~Calderara, and R.~Cucchiara,
``Input Perturbation Reduces Exposure Bias in Diffusion Models,''
\emph{arXiv preprint arXiv:2301.11706}, 2023.

\bibitem{shoemake1985animating}
K. Shoemake,
``Animating rotation with quaternion curves,''
in \textit{Proceedings of the 12th annual conference on Computer Graphics and Interactive Techniques}, 
pp. 245--254, 1985.
%% 39

\bibitem{gao2022data}
Y. Gao, M. Zhou, D. Liu, Z. Yan, S. Zhang, and D. N. Metaxas,
``A data-scalable transformer for medical image segmentation: architecture, model efficiency, and benchmark,''
\textit{arXiv preprint arXiv:2203.00131}, 2022.

\bibitem{petersen2015uk}
S. E. Petersen~\emph{et al.},
``UK Biobank’s cardiovascular magnetic resonance protocol,''
\textit{JCMR}, pp. 1--7, 2015.

\bibitem{isola2017image}
P. Isola, J.-Y. Zhu, T. Zhou, and A. A. Efros,
``Image-to-image translation with conditional adversarial networks,''
in \textit{Proceedings of the IEEE Conference on Computer Vision and Pattern Recognition}, 
pp. 1125--1134, 2017.

\bibitem{kingma2014adam}
D. P. Kingma,
``Adam: A method for stochastic optimization,''
\textit{arXiv preprint arXiv:1412.6980}, 2014.

\bibitem{he2016deep}
K.~He, X.~Zhang, S.~Ren, and J.~Sun,
``Deep residual learning for image recognition,''
in \emph{Proc. IEEE Conf. Comput. Vis. Pattern Recognit. (CVPR)}, 2016, pp. 770--778.

%%%%%%%% 



%%%%%%%%



\bibitem{oring2021autoencoder}
A.~Oring,
``Autoencoder Image Interpolation by Shaping the Latent Space,''
Master's thesis, Reichman University, Israel, 2021.


\bibitem{konz2024anatomically}
N.~Konz, Y.~Chen, H.~Dong, and M.~A.~Mazurowski,
``Anatomically-Controllable Medical Image Generation with Segmentation-Guided Diffusion Models,''
in \emph{International Conference on Medical Image Computing and Computer-Assisted Intervention (MICCAI)}. Springer, 2024, pp.~88--98.


\end{thebibliography}
\end{document}